\documentclass[11pt]{article}
\usepackage{makeidx}  
\usepackage{amsmath,amscd,amsfonts,amssymb,amsthm}
\usepackage[margin=1.25in]{geometry}
\usepackage{algorithmic}
\usepackage{algorithm}
\usepackage[noblocks]{authblk}

\usepackage{tikz}
\usepackage{tikz-cd}
\usepackage{verbatim}
\usepackage{relsize}
\usepackage{url}

\usetikzlibrary{trees,snakes}
\tikzset{fontscale/.style = {font=\relsize{#1}}}

\newcommand{\mypar}[1]{\paragraph{#1}}
\newcommand{\Z}{\mathbb{Z}} 
\newcommand{\Q}{\mathbb{Q}} 
\newcommand{\R}{\mathbb{R}} 
\newcommand{\C}{\mathbb{C}} 
\newcommand{\p}{\mathfrak{p}} 

\newcommand{\Cl}{\operatorname{Cl}}

\newcommand{\OO}{\mathcal{O}}
\newcommand{\OOO}{\underline{\mathcal{O}}}
\newcommand{\OK}{{\mathcal{O}_K}}

\newcommand{\ag}{\mathfrak{a}} 
\newcommand{\bg}{\mathfrak{b}} 
\newcommand{\pg}{\mathfrak{p}} 
\newcommand{\qg}{\mathfrak{q}} 
\newcommand{\Nm}{\mathcal{N}}
\newcommand{\Tm}{\mathcal{T}}

\newcommand{\dist}{\operatorname{dist}}

\newcommand{\diag}{\operatorname{diag}}
\newcommand{\Log}{\operatorname{Log}}
\newcommand{\Vol}{\operatorname{Vol}}

\newcommand{\vect}[1]{\textbf{#1}}

\newtheorem{theorem}{Theorem}
\newtheorem*{theorem*}{Theorem}
\newtheorem{corollary}{Corollary}
\newtheorem{Definition}{Definition}
\newtheorem{proposition}{Proposition}
\newtheorem*{goal*}{Goal}
\newtheorem*{notation*}{Notation}

\newcommand{\cgc}{\text{\sf CGP}}
\newcommand{\pip}{\text{\sf PIP}}
\newcommand{\hsp}{\text{HSP}}

\newcommand{\uset}[1]{\ensuremath{U}(#1)}
\newcommand{\su}[1]{\ensuremath{S}_{#1}\text{-units}}
\newcommand{\sus}{\ensuremath{S}\text{-units}}

\def\cS{\mathcal{S}} 
\def\cL{\mathcal{L}} 
\def\bR{\mathbb{R}}
\def\bZ{\mathbb{Z}}

\def\aisub{\Lambda} 

\newcommand{\ket}[1]{|#1\rangle}
\newcommand{\str}[1]{\mathrm{str}_\nu(#1)} 
\newcommand{\strn}[1]{\mathrm{str}_{\nu,n}(#1)} 
\newcommand{\inp}[2]{\langle #1 | #2\rangle} 
\newcommand{\onorm}[1]{\|#1\|_\infty} 
\newcommand{\fnorm}[1]{\|#1\|_2} 
\newcommand{\vnorm}[1]{\|#1\|}

\newtheorem{fact}{Fact}
\newtheorem{lemma}{Lemma}
\newtheorem{claim}{Claim}
\begin{document}
%

%
%
\title{An efficient quantum algorithm for computing $S$-units and its applications}
%
%

\author[1]{Jean-Fran\c{c}ois Biasse}
\author[2]{Fang Song}

\affil[1]{University of South Florida}
\affil[2]{Portland State University}
\date{}
\maketitle             
\begin{abstract}
\normalsize
In this paper, we provide details on the proofs of the quantum polynomial time algorithm of Biasse and Song (SODA 16) for computing the $S$-unit group of a number field. This algorithm directly implies polynomial time methods to calculate class groups, $S$-class groups, relative class group and the unit group, ray class groups, solve the principal ideal problem, solve certain norm equations, and decompose ideal classes in the ideal class group. Additionally, combined with a result of Cramer, Ducas, Peikert and Regev (Eurocrypt 2016), the resolution of the principal ideal problem allows one to find short generators of a principal ideal. Likewise, methods due to Cramer, Ducas and Wesolowski (Eurocrypt 2017) use the resolution of the principal ideal problem and the decomposition of ideal classes to find so-called ``mildly short vectors'' in ideal lattices of cyclotomic fields. 

\end{abstract}

\section{Introduction}

Let $K$ be a number field of degree $n$ and $\OO$ be an
order
 in $K$ with discriminant $\Delta$. The set of elements $\alpha\in K$ such that 
$\exists (e_i)_{i\leq |S|}\in \Z^ {|S|},\ \ (\alpha) = \pg^ {e_1}\cdots \pg^{e_{|S|}}$ 
is a multiplicative group called the $S$-unit group of $K$. This notion generalizes 
the units of $\OO$ which are $S$-units for $S=\varnothing$, and computing the $S$-unit group 
is an important task in computational number theory. Most notably it applies to the computation of the ideal class group of $\OO$, the resolution of the principal ideal problem in $\OO$, and the resolution of norm equations of the form $\Nm_{L/K}(x) = \theta$ where $\theta\in K$, 
as shown by Simon~\cite{Simon} and Fieker~\cite{Fieker_phd,Fieker}.

 The 
ideal class group $\Cl(\OO)$ is the finite abelian group consisting of the invertible 
fractional ideals of $\OO$ up to principal factors and has order $|\Delta|^{O(1)}$. 
Computing the ideal class group is an essential task in number 
theory that occurs in particular in the resolution of unproven heuristics such as the Cohen-Lenstra 
heuristics \cite{lenstra_heuristic} on class groups of quadratic number field, 
Littlewood's bounds \cite{littlewood} on $L(1,\chi)$, or Bach's bound \cite{bach} on the maximum norm 
of the generators required to generate the class group. Besides being a fundamental problem, 
computing the ideal class group is also strongly connected to number theoretic problems 
occurring in cryptography. For example, it is at the heart of the only known unconditional classical 
subexponential algorithm for integer factorization~\cite{Lenstra_pomerance}. Finding 
relations between elements in $\Cl(\OO)$ also occurs in curve-based cryptography. Indeed, 
both classical~\cite{BCL,jao} and quantum~\cite{jao_souk_quantum} 
subexponential methods for computing isogenies between elliptic 
curves  depend on it.

Given an ideal $\ag\subseteq\OO$, deciding whether or not $\ag$ is principal, 
and if so, finding $\alpha\in\OO$ such that $\ag = (\alpha)$ is called the Principal 
Ideal Problem. It has direct applications to the computation of relative class groups and unit groups, 
and computing the $S$-class group of a number field. It is is also relevant to lattice-based cryptography, which 
has received a considerable attention since it allows quantum-safe cryptosystems and homomorphic 
encryption schemes. For efficiency reasons, there have been many proposals of schemes using lattices  
arising from ideals in the ring of integers of a number field, and in particular principal ideals 
generated by a small element (for example, see the homomorphic encryption 
scheme of Smart and Vercauteren~\cite{fre_smart} and the multilinear maps of 
Garg, Gentry and Halevi~\cite{mult_maps}). It was subsequently proved that solving the 
principal ideal problem in polynomial time directly induces a polynomial time attack on 
schemes relying on the hardness of finding the short generator of a principal ideal~\cite{CDPR15}.

\paragraph{Previous work} Computing the ideal class group and the unit group is a problem that has been 
extensively studied in both the classical and quantum setting. 
Despite these efforts, there are no known polynomial time algorithms for these tasks. 
On the other hand, there are quantum polynomial time algorithms for several hard computational problems in number theory based on 
quantum algorithms for the Hidden Subgroup Problem (HSP). Shor showed that integer 
factorization and the discrete logarithm problem could be solved in polynomial time~\cite{Sho97}, 
and Hallgren described a polynomial time algorithm for solving 
the Pell's equation~\cite{Hal07}. Similar methods were used to compute 
the class group and the unit group in polynomial time in classes of number fields of 
fixed degree~\cite{Hal05,SV05}. The approach of~\cite{Hal05} relies on 
the resolution of the HSP in a bounded and discretized approximation of $\R^m$, 
which does not seem to apply when the degree of the fields grows to infinity. 
In a recent breakthrough, Eisentr\"{a}ger, Hallgren, 
Kitaev and Song~\cite{STOC2014} described a polynomial time algorithm for 
computing the unit group in classes of number fields of arbitrary degree. 
One of the main tools they developed is a continuous HSP definition on $\R^m$ and an efficient quantum algorithm solving it. 
In essence, their new HSP definition enforces stringent \emph{continuity} properties on 
the function that hides the subgroup. This makes the function more amenable to quantum Fourier sampling.

\paragraph{Our contribution} In this paper, we present a quantum algorithm to compute the 
$S$ unit group of a number field of arbitrary degree in polynomial time. It readily applies to the computation of the ideal class group and to the resolution of the principal ideal problem, and well as to other related tasks in computational number theory. 
 We follow a different framework than the previous work in 
constant-degree number fields due to Hallgren~\cite{Hal05}. 
We show that both the ideal class group computation and PIP reduce to a more general 
problem of computing the $S$-unit group for suitable set of prime ideals $S$. 
For example, for the ideal class group computation  $S$ is chosen to
be a succinct generating set of $\Cl(\OO)$. Then we give an efficient
quantum algorithm for computing the $S$-unit group by extending the
work by Eisentr\"{a}ger, Hallgren, Kitaev and Song~\cite{STOC2014}. We
show an efficient quantum reduction from the $S$-unit group problem to
HSP on $\R^m$ as defined in~\cite{STOC2014}, which then can be solved efficiently by the quantum HSP algorithm in~\cite{STOC2014}. 
We also show how to get exact compact representations of the desired
field elements with respect to a given integral basis for $\OO$, while~\cite{STOC2014} 
only returns fixed point rational approximations of the units. Compact 
representations are usually easier for further algebraic processing. Our 
main results are summarized in the next theorem.

\begin{theorem}[$S$-unit group computation]
There is a quantum algorithm for computing the $S$-unit group of a number field $K$ in compact 
representation which runs in polynomial 
time in the parameters $n=\deg(K)$, $\log(|\Delta|)$, $|S|$ and $\max_{\pg\in S}\{\log(\Nm(\pg))\}$,  
where $\Delta$ is the discriminant of the ring of integers of $K$. 
\end{theorem}

\begin{corollary}
    There are quantum polynomial time algorithms for the resolution of the following tasks in computational number theory: 
    \begin{itemize}
        \item Ideal class group computation (under GRH),
        \item $S$-class group computation (under GRH),
        \item Relative class group and unit group computation (under GRH),
        \item Ideal class decomposition in the ideal class group (under GRH),
        \item Principal Ideal Problem,
        \item Ray class group computation (under GRH),
        \item Norm equation resolution,
    \end{itemize}
    where GRH denotes the Generalized Rieman Hypothesis.
\end{corollary}

As an important corollary, combining recent works 
in lattice cryptanalysis~\cite{SOLILOQUY,CDPR15}, our results induce a quantum 
polynomial-time attack on an entire family of cryptosystems relying on the hardness 
of finding a short generator of a principal ideal. 

\paragraph{Response to a recent preprint from de Boer and Felderhoff~\cite{BoerF25}}

In a preprint published on Oct. 3rd 2025 and updated on Oct. 22nd 2025, K. de Boer and J. Felderhoff presented a similar quantum algorithm as our 2016 work~\cite{BiasseS16}, which combined with~\cite{STOC2014} allows one to compute the $S$-unit group of a number field in quantum polynomial time. The work of de Boer and Felderhoff~\cite{BoerF25} uses the same strategy as our previous work~\cite{BiasseS16} with some technical differences. They claimed that the detailed analysis of the degrees of the polynomial dependencies in the work of Biasse-Song~\cite{BiasseS16} (which itself relies on Eisentraeger et al.~\cite{STOC2014,EHKS14long}) was ``not currently possible''. They additionally suggested that the $S$-unit algorithm of~\cite{BiasseS16} did not run in polynomial time~\cite[p. 5]{BoerF25}. Finally, de Boer and Felderhoff claimed that the key differences between their algorithm offered an advantage compared to the stategy of~\cite{BiasseS16}. Below, we comment on these statements.

First and foremost, we would like to confirm that the 2016 paper of Biasse and Song to compute $S$-unit groups~\cite{BiasseS16} does indeed run in polynomial time. There is an unfortunate typo in the published version of~\cite[Th. 5.1]{BiasseS16} which was copied to an earlier version of Theorem~\ref{th:classical-oracle_cost} of this document. It showed a dependency that appeared to be polynomial in the bound of some input coefficients instead of being polynomial in their bit size. The former obviously does not qualify as a polynomial dependency in the size of the input. We apologize for this confusion. Note that the rest of~\cite{BiasseS16}, as well as the previous version of this document remained consistent with a dependency in the bit size (the paragraph above Theorem~\ref{th:classical-oracle_cost} actually made a claim of polynomial behavior in the bit size of the input even in older versions of this document). It is achieved through an ideal exponentiation strategy devised by Eisentraeger et al.~\cite{STOC2014}, and used by Biasse-Song~\cite{BiasseS16}. In this document, it is explained in details in Section~\ref{subsec:comp_split}. While we are sorry that such a confusing typo got included in the published version of our 2016 work~\cite{BiasseS16}, we would like to point out that we had multiple email exchanges with de Boer and Felderhoff during the month preceding the release of their preprint. They never mentioned their concerns regarding the run time of our algorithm, and chose to publish claims of the exponential complexity without seeking our input.

Second, we would like to state that prior work that established the quantum polynomial time of the unit and $S$-unit quantum algorithm were published at STOC 2014~\cite{STOC2014} and SODA 2016~\cite{BiasseS16} respectively. Due to the strict page limits of these venues, the choice was made to not specify the exact degrees of the polynomial dependencies in the input parameters of these algorithms. However, this does not mean that it ``not currently possible'', or that it is even technically challenging. It is a tedious but straightforward process that can be achieved with known methods. We added such analysis in this updated draft. It mostly relies on techniques to control the precision of LLL reductions on fixed points approximations of vectors originally described in 1987 by buchmann and Pohst~\cite{BP87}, and later adapted by Eisentrager et al.~\cite{STOC2014}.

Finally, we comment on the advantage offered by the modified $S$-unit oracle of de Boer and Federhoff. Their main proposal is to precompute some of these powers in ideal multiplication and exponentiation. However, the savings obtained are moderate, and actually turn into a loss if we focus on cases of interest to the cryptography community (for example when $|S|=1$, which is a case that allows the resolution of the Principal Ideal Problem). Indeed, while fewer LLL reductions are required in the oracle described in~\cite{BoerF25}, the size of the entries involved is significantly larger. Indeed, our quantum oracle always performs multiplications between ideals of norm $1$ (following a method already presented in Eisentraeger et al.~\cite{STOC2014}). On the other hand, in~\cite{BoerF25}, the norm of the ideals is non trivial and grows over the course of the algorithm. In the end, the dependency in the degree $n$ in the bit size $\beta$ of the vectors given to the LLL solver is in $n^7$ for~\cite{BoerF25} whereas it is only $n^5$ for our method. Then, the gate count of the LLL resolution has a dependency in $\beta^{3.5}$, which induces a penalty of $n^7$. Since on the other hand our method requires more LLL resolutions over larger matrices, the final dependency in $n$ of the gate count of the oracle described by de Boer and Federhoff is in $n^{31.5}$, while ours is in $n^{35.5}$. However, this slight advantage vanishes once we account for the additional restriction in the work of de Boer and Federhoff that consists in assuming that the set of primes $S$ generates the ideal class group of the field. If $S$ is too small, they suggest enlarging it so that it contains all primes of norm up to $12\log^2|\Delta|$ where $\Delta$ is the discriminant of the field (under the Generalized Rieman Hypothesis, such a set of primes generate the ideal class group~\cite{bach}). The apparent issue of this workaround is that in the important case of the resolution of the Principal Ideal Problem (which can be reduced in polynomial time to the search of $S$-units where $|S|=1$), this induces a penalty of $|S|^{17.5} \in O\left((\log|\Delta|)^{35}\right)$ in the complexity of the gate cost. In cyclotomic fields, where $\log|\Delta|\in\Theta(n\log(x))$, this means that the degree of the highest term in the degree becomes $n^{49}$, which is significantly larger than the $n^{35.5}$ dependency of the gate count of our quantum oracle.       

\section{Technical background}\label{sec:NT}

In this section we review some useful background in number theory and
introduce some definitions and notations. 
The notions of ideal class group and $S$-unit group are standard, and can be found 
in many books. We suggest Neukirch's book~\cite{neukirch} for the fundamental aspects 
of this theory and 
Cohen's book~\cite{cohen} for the algorithmic aspects. We invite the reader who is already 
familiar to these topics to pay attention to the non-standard notion of $E$-ideal that
we introduce in the following. 

\subsection{Number Theory}

\mypar{Number fields} A number field $K$ is a finite extension of $\Q$. Its ring of integers $\OK$ has the structure 
of a $\Z$-lattice of degree $n=[K:\Q]$, and the orders $\OO\subseteq\OK$ are the sublattices of $\OK$ 
which have degree $n$ and which are equipped with a ring structure. Throughout this paper, we assume that 
$\OO$ is an order in a number field $K$, and we denote by $\omega_1,\ldots,\omega_n$ a $\Z$-basis, 
that is $\OO = \Z\omega_1\oplus\ldots\oplus\Z\omega_n.$ 
A number field has $n_1$ real embeddings and $n_2$ pairs of complex embeddings which we denote 
$(\sigma_j:K\to \R)_{j\leq n_1},((\sigma_j,\overline{\sigma_j}):K\to
\C)_{j\leq n_2}$ with $n_1+n_2 = n = \mathrm{deg}(K)$. 
These embeddings define two essential maps, namely the norm and trace maps which are given by 
$\Tm(x):=\sum_\sigma \sigma(x)\in\Q$ and $\Nm(x):= \prod_\sigma \sigma(x)\in\Q$.
The trace map is additive while the norm map is multiplicative. Note that $\Tm(\OO)\subseteq\Z$ and 
$\Nm(\OO)\subseteq\Z$. We measure the size of the ring $\OO$ by
$\log|\Delta|$ where $\Delta := \left(
  \det(\sigma_j(\omega_k))\right)^2$ is its discriminant, and it
equals the volume of the fundamental domain of $\OO$. 
Equivalently, the discriminant can be defined from the trace map by $\Delta:= \det(\Tm(\omega_i\omega_j))_{i,j\leq n}$.

\mypar{The ideal class group} The fractional ideals of $\OO$ generalize the notion of ring ideals of $\OO$. They are the subsets of $K$ of the 
form $\ag = \frac{1}{d}I$ where $d\in\Z^+$ and $I\subseteq \OO$ is an (integral) ideal of $\OO$. A fractional ideal $\ag$ is 
invertible if $\ag^{-1} := \{ x \in K \mid \ x\ag\subseteq \OO\}$ is also a fractional ideal. 
The invertible fractional ideals have a multiplicative group structure, and the principal fractional ideals are 
one of its subgroups. The ideal class group is defined by 
$$\Cl(\OO) := \mathcal{I}/\mathcal{P},$$
where $\mathcal{I}$ is the multiplicative group of fractional invertible ideals of $\OO$ and 
$\mathcal{P}$ is the subgroup of elements of $\mathcal{I}$ that are principal. 
This means that we identify $\ag$ and $\bg$ in $\Cl(\OO)$ if there is
$\alpha\in K$ such that $\ag=(\alpha)\bg$. 
Ideals are sublattices of $\OO$ of rank $n$, and we define their norm by $\Nm(I):= \left| \OO / I\right|$. This notion 
naturally extends to fractional ideals using the multiplicative rule $\Nm(\ag/\bg) := \Nm(\ag)/\Nm(\bg)$. This 
notion of norm extends the norm on $K$ in the sense that if $\ag = (\alpha)$, then $\Nm(\ag)=\Nm(\alpha)$.

\mypar{The $S$-unit group} The $S$-units are a generalization of the units $\OO^*$, which are the invertible elements of $\OO$. The unit 
group can alternatively be defined as the $\alpha\in\OO$ with $|\Nm(\alpha)|=1$, or the $\alpha\in\OO$ such 
that $(\alpha) = \OO$. The unit group $\OO^*$ satisfies 
$\OO^* \simeq \mu\times \langle \varepsilon_1\rangle \times \ldots \times \langle \varepsilon_r\rangle,$ 
where $r:= n_1 + n_2 - 1$, $\mu$ is the set of roots of unity and the $\varepsilon_i$ are torsion-free 
units. Let $S = \{\p_i\}$ be a finite set of prime ideals of $\OO$, the $S$-units are the elements 
$\alpha\in K$ such that there is $ (v_i(\alpha))_{i\leq |S|}\in \Z^ {|S|}$ with $(\alpha) = \pg_1^ {v_1(\alpha)}\cdots \pg_{|S|}^ {v_{|S|}(\alpha)}.$ 
Note that the $S$-units are elements of $K$. They form a multiplicative group $\uset{S}$ satisfying 
$\uset{S} \simeq \mu\times \langle \varepsilon_1\rangle \times \ldots \times \langle \varepsilon_{r+|S|}\rangle,$ 
where $r:= n_1 + n_2 - 1$, $\mu$ is the set of roots of unity and the $\varepsilon_i$ are torsion-free 
$S$-units. 

\mypar{$E$-ideals} The number field $K$ can be naturally embedded into
$E:=\R^{n_1}\times\C^{n_2}$ by setting $z\in \OO \mapsto(\sigma_1(z),\ldots,\sigma_{n_1+n_2}(z))$. As in~\cite{STOC2014}, we  
denote by $\OOO$ the image of $\OO$ via this embedding. The set $\OOO$ inherits from the lattice 
structure of $\OO$, i.e. it can be identified as a lattice in $\R^n$, as well as from the multiplication between 
elements (which is performed component-wise). 
The image of the fractional ideals of $K$ in $E$ are lattices $\Lambda\subseteq E$ with the property that 
$x\Lambda\subseteq\Lambda$ for all $x\in\OOO$. We define the $E$-ideals as all the lattices in $E$ satisfying 
this property. When there is no ambiguity, we identify a fractional ideal of $\OO$ and the corresponding 
$E$-ideal.

\begin{Definition}[$E$-ideals]
Let $E:=\R^{n_1}\times\C^{n_2}$ and $\OOO$ the image of $\OO$ via the embedding $K\rightarrow E$. An 
$E$-ideal is a lattice $\Lambda\subseteq E$ such that 
$\forall x\in \OOO, \ x\Lambda\subseteq\Lambda.$
\end{Definition}

\subsection{HSP resolution}

\mypar{Continuous HSP} We review the definition of continuous HSP
proposed by Eisentr\"{a}ger et al.~\cite{STOC2014}, for which they have
shown an efficient quantum algorithm.  
\begin{Definition}[Continuous HSP over $\R^m$]\label{def_HSPRm}
The unknown subgroup $L\subseteq \R^m$ is a full-rank lattice satisfying some promise: the norm of the shortest vector is at least $\lambda$ and the unit cell volume is at most $d$. The oracle has parameters $(a,r,\varepsilon)$. Let $f: \R^m \rightarrow \mathcal S$ be a function, where $\mathcal S$ is the set of unit vectors in some Hilbert space. We assume that $f$ hides $L$ in the following way.
\begin{enumerate}
  \item\; $f$ is periodic on $L$, i.e. $f(x)=f(x+v)$ for all $x\in \R^m$ and $v\in L$;
  \item\; $f$ is Lipschitz with constant $a$, i.e. $\bigl\|\ket{f(x)}-\ket{f(y)}\bigr\| \le a\|x-y\|$ for all $x,y\in \R^m$;
  \item If the distance between the cosets ($x\bmod L$) and ($y\bmod L$) is greater or equal to $r$, i.e.\ if $\min_{v\in L}\|x-y-v\|\ge r$, then $\bigl|\langle{f(x)}|{f(y)}\rangle\bigr|\le\varepsilon$.
\end{enumerate}
Under these conditions, the problem is to compute a basis of $L$ by a quantum algorithm that can make oracle calls $\ket{x}\mapsto\ket{x}\otimes\ket{f(x)}$.
\end{Definition}
Actually, the definition also applies more generally to other topological
groups $G = \R^k/\Lambda \times D$ with a proper metric on
$G$~\cite[Sect.6.1]{STOC2014}. Here $G$ is decomposed to a continuous
part, which is the quotient of $\R^k$ over some lattice $\Lambda$, and
a discrete part that is finitely generated. It is nonetheless
sufficient to consider HSP on $\R^m$, because the more general case can be reduced to HSP
on $\R^m$~\cite{STOC2014}, and hence can be solved efficiently. In the following, we define a control group $G$ on which a first version of our HSP oracle will be defined. We prove HSP properties on $G$, and then extend it to $\R^m$.

Suppose $\sigma_1,\ldots,\sigma_{n_1}$ are the real embeddings of $K$, and that
$\sigma_{n_1+1},\ldots,\sigma_{n_1+n_2}$ are the (non-pairwise conjugate) complex embeddings of $K$. Assume also that $S = \{\pg_1,\ldots,\pg_s\}$ where $\Nm(\pg_i) = p_i^{e_1}$. An element $x\in U_S$ satisfies $\prod_{i=1}^{n_1+2n_2} \sigma_i(x) = \Nm(x) = \prod_i p_i^{e_iv_i(x)}$. This means that we know that 
$$
\log|\sigma_{1}(x)| = -\sum_{i=2}^{n_1}\log|\sigma_i(x)| - 2\sum_{i=n_1+1}^{n_2}\log|\sigma_i(x)| + \sum_{i\leq s} e_iv_i(x)\log p_i.
$$
Therefore, $x\in U_S$ corresponding to $(x_1,\ldots,x_n)=(\sigma_1(x),\ldots,\sigma_n(x))\in\R^{n_1}\times\C^{n_2}$ is uniquely identified by the element $x^G\in G := \R^{n_1 + n_2 - 1}\times \Z_2^{n_1} \times \left(\R/\Z\right)^{n_2}\times \Z^s$ where
\begin{itemize}
    \item $x_i^G = \log(|x_{i+1}|)$ for $1\leq i< n_1+n_2$,
    \item $x_i^{G} = \delta_i\in\Z_2$ where $x_{i-n_1-n_2+1} = (-1)^{\delta_i}|x_{i-n_1-n_2+1}|$ for $n_1+n_2\leq i< 2n_1 + n_2$.
    \item $x_i^{G} = \theta_i\in\R/\Z$ where $x_{i-2n_1-n_2+1} = e^{2i\pi\theta_i}|x_{i-2n_1-n_2+1}|$ for $2n_1+n_2\leq i< 2n_1 + 2n_2$.
    \item $x_i^G = v_{\pg_{i-2n_1 + 2n_2}}(x)$ for $2n_1+2n_2\leq i< 2n_1+2n_2+s$.
\end{itemize}
Conversely, we have a map $\phi:G\rightarrow \R^{n_1}\times\C^{n_2}$ such that $\phi(x^G)=x$ by choosing $|x_i| = e^{x_{i+1}^{G}}$ for $1\leq i<n_1+n_2$, and 
\begin{equation}\label{eq:last-coordinate}
    |x_1| = \frac{\prod_{i=1}^{s}p_{i}^{e_ix_{i+2n_1+2n_2}^G }}{\prod_{i=2}^{n_1}|x_i|\prod_{i=n_1+1}^{n_1+n_2}|x_i|^2}.
\end{equation}
Then we do 
\begin{itemize}
    \item $x_i\leftarrow (-1)^{x^G_{i+x_1+x_2-1}}|x_i|$ for $1\leq i\leq n_1$ and
    \item  $x_i\leftarrow e^{2i\pi x^G_{i + 2x_1+x_2 - 1}}|x_i|$ for $n_1 < i \leq n_1+n_2$.
\end{itemize}
\begin{Definition}[Control group $G$]
Let $K$ be a number field of signature $(n_1,n_2)$, and $S$ a set of primes above $(p_i)_{i\leq s}$. 
We define the following groups: 
\begin{itemize}
 \item $G = \R^{n_1 + n_2 - 1}\times \Z_2^{n_1} \times \left(\R/\Z\right)^{n_2}\times \Z^s$ the \emph{control group}.
 \item $L = U_S^G\subseteq G$ the image of the $S$-unit group of $K$, which is a lattice. 
\end{itemize}
\end{Definition}
The map $\phi$ is readily extended beyond elements of $G$ that correspond to an $S$-unit. In this case, $\phi(u,v)\in\R^{n_1}\times \C^{n_2}$ for $u\in\R^{n_1+n_2-1}\times \Z_2^{n_1}\times (\R/\Z)^{n_2}$ and $v\in\Z^s$ does not necessarily correspond to an element $x\in K$ with $\Nm(x) = \prod_i p_i^{e_iv_i}$. On the other hand, in general, there is no canonical way to map an element of $\R^{n_1}\times \C^{n_2}$ that is not an $S$-unit to an element of $G$.

The control group can be seen as the projection of $\tilde{G} = \R^k\times \Z^l$ where 
\begin{itemize}
 \item $k = n_1 + 2n_2 -1$.
 \item $l = n_1 + s$.
\end{itemize}
We denote by $\gamma:\tilde{G}\rightarrow G$ the projection map, and by $\tilde{L}\subseteq\tilde{G}$ the 
pre-image of $L$ by $\gamma$. It is a lattice in $\tilde{G}$. We also construct an 
oracle $g = f_q\circ f_c: G/L\rightarrow \mathcal{H}$ where 
\begin{itemize}
 \item $f_c(t,v) = e^{\textbf{t}}\underline{\OO}\prod_{\pg\in S}\pg^{-v_i}$, which is a lattice in $E:= \R^{n_1}\times \C^{n_2}$.  
 \item $f_q(L_{E}) = \ket{L_E}: = \gamma \sum_{v\in L_E} g_s(v) \ket{\strn{v}}$ which is a quantum state (see Section~\ref{sec:quantum-encoding} for a definition of the straddle encoding $\ket{\strn{v}}$ instroduced in~\cite{STOC2014}.
\end{itemize}

To prove the HSP properties of $f: G\rightarrow \mathcal{H}$, we 
need a notion of distance between ideals of $E = \R^{n_1}\times \C^{n_2}$. An ideal in $E$ is a 
lattice that is stable by multiplication by elements in $\underline{\OO}$ (the embedding of $\OO$ in $E$). 
We deal with elements in $E$ by embedding them in $\R^n = \R^{n_1 + 2n_2}$ (via $z\in\C\mapsto \mathfrak{Re}(z),\mathfrak{Im}(z)$). 
Each $E$-ideal $\mathcal{L}$ can be defined by a matrix $M_\mathcal{L}\in\R^{n\times n}$ whose rows are a 
$\Z$-basis of $\mathcal{L}$. Note that $E$-ideals $\mathcal{L},\mathcal{L'}$ can be multiplied, but $M_{\mathcal{L}\mathcal{L'}}$ 
is in general \emph{not} equal to $M_\mathcal{L}M_\mathcal{L'}$. 

\begin{notation*}
The Euclidean norm is used in different spaces. When there is a potential ambiguity, we use a subscript 
to specify the space. More specifically, suppose there is a group $H$ and $s,t$ such that 
$\alpha:H\hookrightarrow \R^s\times \C^t$, then for $x\in H$, 
we denote by $\|x\|_H = \|\alpha(x)\|$, i.e. if 
$x = (x_1,\ldots,x_{s+t})$, then $\|x\|_H = \sqrt{\sum_{i\leq s} |x_i|^2 + 2\sum_{i> s} |x_i|^2}$. 
\end{notation*}

\begin{Definition}[Matrix distance between $E$-ideals]
Let $\mathcal{L},\mathcal{L}'$ be two $E$-ideals. 
We define the \emph{matrix distance} between $\mathcal{L}$ and $\mathcal{L}'$ by 
$$
\dist(\cL,\cL') = \inf_{A,M_\cL,M_{\cL'}}\{\|A\|_2:\ M_\cL = M_{\cL'}e^A,\ A\in \operatorname{Gl}_n(\R)\}
$$
\end{Definition}

As in~\cite{EHKS14long}, given an element $x\in E$, we define the 
matrix $\diag(x)\in\R^{n\times n}$ which is \emph{not} exactly a diagonal matrix. 
$$\diag(x) := 
\begin{pmatrix}
x_1 &        &         &                &        &             \\
    & \ddots &         &                &        &             \\
    &        & x_{n_1} &                &        &             \\
    &        &         & \Xi(x_{n_1+1}) &        &             \\
    &        &         &                & \ddots &             \\
    &        &         &                &        & \Xi(x_{n_2}) 
\end{pmatrix}\text{ where   }
\Xi(z) := 
\begin{pmatrix}
 \mathfrak{Re}(z) & -\mathfrak{Im}(z)\\
 \mathfrak{Im}(z) & \mathfrak{Re}(z)
\end{pmatrix}.
$$
Given $x\in E$, the above matrix has the important property that $M_{(x)\cdot\mathcal{L}} = M_\mathcal{L}\cdot\diag(x)$ where 
$\mathcal{L}$ is an $E$-ideal, and $(x)$ denotes the $E$-ideal $x\cdot \underline{\OO}$ (a principal ideal generated by $x$). This 
is a case where ideal multiplication corresponds to a product of matrices (although $\diag(x)$ is not $M_{(x)}$). 
\begin{lemma}\label{lem:diag-matrices}
Matrices of the form $\diag(x)$ have the following properties: 
\begin{enumerate}
 \item $\forall x_1,x_2\in E$, $\diag(x_1) + \diag(x_2) = \diag(x_1 + x_2)$.
 \item $\forall x_1,x_2\in E$, $\diag(x_1)\cdot\diag(x_2) = \diag(x_1\cdot x_2)$.
 \item $\forall x\in E$, $e^{\diag(x)} = \diag(e^x)$ where $e^x = (e^{x_1},\ldots,e^{x_{n_1+n_2}})$.
 \item $\forall x\in E$, if $\|\diag(x)-I\|<1$, then $\log(\diag(x)) = \diag(\log(x))$ where 
 $\log(x) = (\log(x_1),\ldots,\log(x_{n_1+n_2}))$. 
\end{enumerate}
\end{lemma}
\begin{proof}
For 1) and 2), it suffices to check that $\forall z,z'\in \C$, $\Xi(z) + \Xi(z') = \Xi(z + z')$, 
and $\Xi(z)\cdot\Xi(z') = \Xi(zz')$. Then, since $e^A = \sum_{k=0}^\infty \frac{A^k}{k!}$, 
we have 
$$
e^{\diag(x)} = \sum_{k=0}^\infty \frac{\diag(x)^k}{k!} = \diag\left( \sum_{k=0}^\infty \frac{x^k}{k!}\right) = \diag(e^x).
$$
Likewise, to prove 4), we simply use the convergence of $\sum_{k=1}^\infty (-1)^{k+1}\frac{(B-I)^k}{k}$ 
to $\log(B)$ whenever $\|B-I\|<1$. 
\end{proof}

\section{High level overview}\label{sec:high_level}

Our algorithms for the Class Group Problem (\cgc{}) and the Principal Ideal Problem (\pip{}) consist of
reductions to the continuous hidden subgroup problem in two steps, and invoking the quantum HSP
algorithm~\cite{STOC2014} at the end. 
\begin{align*}
 \cgc & \leq_C \su{\cgc} \leq_Q \hsp(\R^{O(n)}), \\
 \pip & \leq_Q \su{\pip} \leq_Q \hsp(\R^{O(n)}) \, .
\end{align*}
Specifically, we first reduce them to
$S$-unit problems with proper choices of $S$, which are almost
entirely \emph{classical} except that we apply a quantum algorithm for
factoring ideals in the case of \pip\footnote{These reductions are 
straightforward. But classical algorithms typically compute the $S$-unit group
by solving \cgc{} and solving instances of \pip{} first. Our quantum algorithm tackles these problems in 
the reverse order.}. 
We describe these reductions to \sus{} problems in Sect.~\ref{sect_red2su}. 
Next we show a \emph{quantum} reduction from \sus{} problem for any
$S$ to $\hsp(\R^m)$, with $m = O(|S|,n)$. This is the main technical contribution of
this work and it generalizes the reduction from
(ordinary) unit-group problem to \hsp{} by Eisentr\"{a}ger et
al.~\cite{STOC2014}. The details will appear in
Section~\ref{sec:applications}, and we give an overview below. 

Given $S=\{\p_1,\ldots,\p_k\}$, we want to
establish a function that hides the $S$-unit group according to
Definition~\ref{def_HSPRm}. To warm up, we review the reduction for the ordinary unit group (i.e.,
$S=\emptyset$)~\cite{STOC2014}. 

\mypar{Review: reduction for unit-group~\cite{STOC2014}} Observe that the unit group can be identified as a subgroup of
$G := \R^{n_1+n_2}\times\Z_2^{n_1}\times(\R/\Z)^{n_2}$, and the mapping 
\begin{align*}
\varphi:\quad& (u_1,\ldots,u_{n_1+n_2},\mu_1,\ldots,\mu_{n_1},\theta_1,\ldots,\theta_{n_2})\\
   \mapsto &(\ldots,(-1)^{\mu_i}e^{u_i},\ldots,\ldots,e^{2\pi i
  \theta_i}e^{u_i},\ldots) \, .
\end{align*}
 translates between the so-called \emph{log coordinates} and the conjugate vector
 representation. To see this, note that under canonical embeddings, any
 $z\in\OO$ has the conjugate vector representation
 $(\ldots,\sigma_i(z),\ldots)\in \R^{n_1}\times\C^{n_2}$. If in addition $z$ is invertible,
 then $\sigma_i(z) \neq 0$. Therefore, we can write $\sigma_i (z) = (-1)^{\mu_i}e^{u_i}$ with $\mu_i \in \Z_2$ and $u_i \in \R$ if $\sigma_i$
 is real, or $ \sigma_i(z) = e^{2\pi i \theta_i}e^{u_i}$ with $\theta_i \in
 \R/\Z$ and $u_i \in \R$ if $\sigma_i$ is
complex. 

Now one defines $f$ in~\cite{STOC2014} as composition of two mappings: 
\[f: G \stackrel{g}{\longrightarrow}  \{E\text{-ideals}\} \stackrel{f_q}{\longrightarrow} \{\text{quantum states}\} \, .\]
Given $x\in G$, $g(x) : = \varphi(x)\OOO \subseteq E$ produces an $E$-ideal which is a transformed lattice of $\OOO$. This is
motivated by the fact that $\alpha \OO =\OO$ for any unit $\alpha\in
\OO^*$. Actually, one can verify easily that $g(x) = g(y)$ \emph{iff.}
$\varphi(x-y) \in \OO^*$. Namely $g$ is periodic on $\OO^*$. For
lacking of a canonical basis to represent real-valued lattices uniquely, which
is needed to apply the quantum HSP algorithm, a quantum mapping $f_q$
follows. It encodes a lattice $L$ into a quantum state $\ket{L}$ that
is roughly composed of quantum superposition over all lattice points, and hence
provides a canonical representation for lattices. We will give more
details of the quantum encoding in Sect.~\ref{ssec:oracle}. 

Very informally, one can show that small shift on an input to $g$ causes
small variance on the output lattice, but two inputs that are far
apart modulo any unit will be mapped to lattices that have small
overlap. Moreover, $f_q$ preserves the ``closeness'' of
lattices. Namely, quantum encodings of two lattices will have
substantial inner product if and only if the lattices are very well lined
up. To formalize these statements and thus proving the HSP properties,
nonetheless, turn out to be highly non-trivial. It involves for
example defining proper distance measures on various input and output
spaces, and analyzing the continuity properties of $f$ with respect to
these metrics. This has been a great amount of efforts in~\cite{STOC2014} with further
details in~\cite{EHKS14long}

Other than these analytic properties, to make an efficient reduction,
one needs to implement $f=f_q\circ g$ efficiently. In fact, $f_q$ can
be implemented efficiently on a quantum computer by standard
techniques. Computing $g$, on the other hand, is much more
tricky. For instance $e^{u_i}$ will involve
doubly-exponential numbers if we manipulate them naively. Instead one splits the
computation into small pieces, in the spirit of repeated squaring, and
carefully controls the precision. There is one
key observation that guarantees that the size of any intermediate step
does not blow up. That is $\Nm(z) = \pm 1$ for any unit $z$ and hence
$\prod_{i=1}^{n_1} e^{u_i} \prod_{j=1}^{n_2} e^{2 u_{n_1+j}} =
1$. This indicates one redundant coordinate, and we can hence
restrict $f$ on
$\R^{n_1+n_2-1}\times\Z_2^{n_1}\times(\R/\Z)^{n_2}$ instead. This
characterization is also essential to show a suitable bound on the volume of the unit cell of $\OO^*$.
 
\newcommand{\sg}{{\hat g}}
\newcommand{\sG}{{\hat G}}
\mypar{Reducing \sus{} to \hsp} It is now easier to describe our
generalized reduction for \sus. Let $S=\{\p_1,\ldots,\p_k\}$. By definition, if $\alpha\in \OO$ is an $S$-unit, we have 
$$\alpha\cdot \OO\cdot \pg_1^{-v_{\pg_1}(\alpha)}\cdots
\pg_{k}^{-v_{\pg_{k}}(\alpha)} = \OO,$$ 
where $v_\pg(\alpha)$ is the coefficient of $\pg$ in the power of $(\alpha)\OO$ (the valuation of $\alpha$ at $\pg$).
Therefore the group of $S$-units $U_S$ corresponds 
to the subgroup of $G = \R^
{n_1+n_2-1}\times \Z_2^{n_1}\times (\R/\Z)^{n_2}{\times \Z^{s}}$ such that $\phi(y,v)\cdot \OOO \cdot
  \pg_1^{-v_1}\cdots \pg_{|S|}^{-v_{|S|}} = \OOO$.  This
motivates us to define the function $f_c: G \to \{E\text{-ideals}\}$
by: 
\[f_c:(y,v_1,\ldots,v_{|S|})\longmapsto \phi(y,v)\cdot \OOO \cdot
  \pg_1^{-v_1}\cdots \pg_{|S|}^{-v_{|S|}}\, .\]

We can show that $\sg$ is periodic on $U_S$. We
then apply the same quantum encoding $f_q$ on the output of
$\sg$. Namely, our oracle function behaves like: 
\[f: G \stackrel{f_c}{\longrightarrow}  \{E\text{-ideals}\} \stackrel{f_q}{\longrightarrow} \{\text{quantum states}\} \, .\]

While the classical mappings $g$ and $f_c$ bear some similar motivation and we reuse
$f_q$, to prove HSP properties of our function $f$ is not 
straightforward. We need to define new metrics tailored to the
specific group structure that the $S$-units belong and the $E$-ideals
(lattices in $\R^n$) that
our $f_c$ may possibly generate. Then we show quantitatively that
under these metrics, small variance in inputs induces slightly
perturbed lattices, whereas large variance of inputs modulo any
$S$-units will induce with high fraction of mismatch. Finally we
relate the new metrics to the analysis of~\cite{STOC2014} and conclude
the \hsp{} properties. We further extend the function $f$ to
obtain an \hsp{} instance on $\R^m$ and work out the necessary bounds
$(\lambda,d)$ as required, which allows us to invoke the quantum HSP
algorithm to recover $U_S$.

\section{Defining the oracle function
  $(\textbf{y},\textbf{v})\mapsto
  |\varphi(\textbf{y}) \underline{\OO}\prod_{\pg\in
    S}\pg^{-v_i}\rangle$}
\label{ssec:oracle}
Our algorithm relies on a classical oracle that takes an element 
in $G$ and maps it to 
$$f_c(y,v_1,\cdots.v_{|S|}) = \phi(y,v)\cdot \OOO \pg_1^{-v_1}\cdots\pg_{|S|}^{-v_{|S|}} \, .$$
Then the corresponding lattice is encoded by an approximation of the superposition 
of all its points denoted by $f_q$. 
As $G = \R^ {n_1+n_2-1}\times \Z_2^{n_1}\times (\R/\Z)^{n_2}\times \Z^{|S|}$, we need to work 
with approximations of real numbers. To perform the necessary arithmetic operations between 
$E$-ideals presented in Section~\ref{subsec:arithmetic}, we use the results of Buchmann and 
Pohst~\cite{BP87} and of Buchmann and Kessler~\cite{BK93} which rely on fixed point 
approximations. More specifically, they use the rounding of the $2$-adic expansion of real 
numbers. The approximation of $a\in\R$ of precision $q\in\Z_{>0}$ is $\widehat{a}\in\Z$ such 
that $\left| \frac{\widehat{a}}{2^q} - a\right| \leq \frac{1}{2^{q+1}}$. However, it seems that 
this notion of approximation is not stable when we multiply two approximate numbers together. 
We made a slight adjustment to their claims to incorporate the case of approximations 
such that $\left| \frac{\widehat{a}}{2^{q_0}} - a\right| \leq \frac{1}{2^{q}}$ for some 
$q_0\geq q$. Then in Section~\ref{subsec:complexity} we show that the classical oracle 
runs in polynomial time with respect to the size of the input. 

\subsection{Splitting up the computation}\label{subsec:comp_split}

Let $(y,v_1,\cdots,v_|S|)\in \R^ {n_1+n_2-1}\times \Z_2^{n_1}\times (\R/\Z)^{n_2}\times \Z^{|S|}$. The naive 
computation of 
$$f_c(y,v_1,\cdots,v_{|S|}) = \phi(y,v)\cdot \OOO \cdot \pg_1^{-v_1}\cdots \pg_{|S|}^{-v_{|S|}}$$
involves computing $(e^{u_i})_{i\leq n_1+n_2}$, where $y = (u_1,\cdots,n_{n_1+n_2},\theta)$ and $u_1$ is computed by the rule given by~\eqref{eq:last-coordinate} with 
a phase $\theta\in\Z_2^{n_1}\times (\R/\Z)^{n_2}$. Any rational approximation of $e^{u_i}$ has at least 
$\left\lceil \log_2(e^{u_i}) \right\rceil \in O(u_i)$ bits where $\log_2$ denotes the base~$2$ logarithm. 
As this is exponential in the bit size of the entry, we need to proceed differently to evaluate $f_c$. The 
authors of~\cite{STOC2014} described a way to split up the computation ensuring that we only manipulate values of 
polynomial size. We adapt this method to our specific classical oracle that differs 
by a term of the form $\prod_{\pg_i\in S}\pg_i^{-v_i}$ from the one described in~\cite{STOC2014}. 

Our input can be split between $(u_1,\cdots,u_{n_1+n_2},v_1,\cdots,v_{|S|})\in \R^{n_1+n_2}\times \Z^{|S|}$ and 
a phase $\theta\in \Z_2^{n_1}\times (\R/\Z)^{n_2}$. As mentioned in~\cite{STOC2014}, the phase can be dealt with 
separately and is not computationally problematic. To make our presentation simpler, we show how to split up the 
computation in the absence of phase. To avoid the expensive computations with the $e^ {u_i}$, we use $E$-ideal 
arithmetic which we analyze in Section~\ref{subsec:arithmetic}. Our main concern when splitting up the computation 
is that we want to reduce it to operations between $E$-ideals of determinant $\sqrt{|\Delta|}$. This gives us 
upper and lower bounds on the vectors in play, which in turns bounds the computational complexity of arithmetic operations 
as we see in Section~\ref{subsec:complexity}. 

Let  $(u_1,\cdots,u_{n_1+n_2},v_1,\cdots,v_{|S|})\in\R^{n_1+n_2}\times\Z^{|S|}$ be an input vector where $u_1$ satisfies the 
condition given by~\eqref{eq:last-coordinate}. We can separate the evaluation of the oracle in two steps by rewriting it as 
$$\left(u_1,\cdots,u_{n_1+n_2-1},u'_{n_1+n_2},0,\cdots,0\right) + \left(0,\cdots,0,\frac{1}{2}\sum_je_jv_j\log(p_j),v_1,\cdots,v_{|S|}\right).$$
where $u'_{n_1+n_2} = -\frac{1}{2}\sum_{j\leq n_1}u_j - \sum_{n_1<j< n_1+n_2}u_j$. The first term is 
evaluated the same way as~\cite{STOC2014}. More specifically, we separate real numbers between integer and fractional part. We define $(r_j)_{j\leq n_1+n_2}\in\Z^{n_1+n_2}$ and 
$(s_j)_{j\leq n_1+n_2}\in [0,1)^{n_1+n_2}$ by $u_j = r_j + s_j$ for $j<n_1+n_2$, 
$r_{n_1+n_2} := -\sum_{j<r_1+r_2}r_j$ and $s_{n_1+n_2} := u'_{n_1+n_2}-r_{n_1+n_2}$. 
As $s_i < 1$, we calculate $e^{s_i}$ to a given precision $q$ 
by using the formula $ e^x = \sum_{k\leq M}\frac{x^ k}{k!} + O(x^{M+1})$. The number of terms in the sum has to satisfy 
$M\in O(q)$. This way, we can compute $\phi(s_1,\cdots,s_{n_1+n_2})=(e^ {s_1},\cdots.e^{s_{n_1+n_2}})$ and the 
corresponding $E$-ideal $A_{-1}:=(e^ {s_1},\cdots.e^{s_{n_1+n_2}})\cdot\OOO$ by multiplication with each generator of $\OOO$. 
Let $(a_k^{(j)})\in\{-1,0,1\}$ be such that $r_j = \sum_{k\leq \lceil\log_2(r_j)\rceil}a_j^{(k)}2^k$ is the binary decomposition of 
 $r_j$ for $j < n_1+n_2$ and 
$a_k^{(n_1+n_2)} := -\sum_{j<n_1+n_2}a_k^{(j)}$ and 
$\log_2(r) := \max_j\lceil\log_2(r_j)\rceil$. Note that we have $u'_{n_1+n_2} = \sum_k a_k^{(n_1+n_2)}2^k$, but 
the $a_k^{(n_1+n_2)}$ are not its binary decomposition. They take values in $[-n_1-n_2,n_1+n_2]$. The 
$E$-ideal generated by the integer part of the $u_i$ satisfies 
\begin{align}
(e^{r_1},\cdots,e^{r_{n_1+n_2}})\cdot\OOO 
& = \prod_{k\leq \log_2(r)}\left( e^{a_1^{(k)}2^k},\cdots,e^{a_{n_1+n_2}^{(k)}2^k}\right)\cdot\OOO\nonumber\\
&= \prod_{k\leq \log_2(r)}\left[ \underbrace{\left( e^{a_1^{(k)}},\cdots,e^{a_{n_1+n_2}^{(k)}}\right)\cdot\OOO}_{A_k}\right] ^{2^k}.\label{eq:def_Aj}
\end{align}
The norm of the $E$-ideals $A_k$ for $k\leq \log_2(r)$ is $\Nm(A_k) = e^{\sum_j a_j^{(k)}}\Nm(\OOO)=1$. Therefore 
$\det(A_k) = \sqrt{|\Delta|}$. 

Likewise, the bit size of $e^{e_1v_i\log(p_i)}$ is at least proportional to $v_i$, and therefore exponential 
in the bit size of $v_i$ which is part of the input. Therefore, we need to split up the computation of the 
$E$-ideal
$$\left(0,\cdots,0,\frac{1}{2}\sum_je_jv_j\log(p_j),v_1,\cdots,v_{|S|}\right)
\longmapsto \left( 1 , \cdots , 1 , e^{\frac{1}{2}\sum_je_jv_j\log(p_j)}\right)\cdot\OOO
\cdot\prod_{j}\pg_j^{-v_j}.$$
Let $(b_j^{(k)})$ such that $v_j = \sum_{k\leq \lceil\log_2(v_j)}b_j^{(k)}2^k$ and 
$\log_2(v):= \max_j\log_2(v_j)$. Then we have the decomposition 
\begin{align}
\left( 1 , \cdots , 1 , e^{\frac{1}{2}\sum_je_jv_j\log(p_j)}\right)\cdot\OOO
\cdot\prod_{j}\pg_j^{-v_j}
& = \prod_{j\leq |S|} \left( \left( 1 , \cdots , 1 , e^{e_j\log(p_j)}\right)\cdot\OOO\cdot\pg_j^{-1}\right)^{v_j}\nonumber\\
&= \prod_{j\leq |S|} \prod_{k\leq \log_2(v)} \left( \left( 1,\cdots,1,e^{e_j\log(p_j)}\right)\cdot\OOO\cdot \pg_j^{-1}\right)^{b_j^{(k)}2^k}.\nonumber\\
&= \prod_{k\leq \log_2(v)}\left( \prod_{j\leq|S|}\left( \underbrace{(1,\cdots,1,p^{e_j})\cdot\OOO\cdot \pg_j^{-1}}_{B_{j,k}}\right)^{b_j^{(k)}}\right)^{2^k}\label{eq:def_Bj}
\end{align}
The calculation is decomposed the following way: first compute $B_k := \prod_{j\leq |S|}B^{b_j^{(k)}}_{j,k}$ which involves 
$\log_2(v)\cdot|S|$ multiplications between the $E$-ideals $B_{j,k}$ which have determinant $\sqrt{|\Delta|}$, and then 
return $\prod_{k\leq \log_2(v)}B_{k}^{2^k}$ which requires at most $\log_2(v)^2$ multiplications between the 
$E$-ideals $B_k$ which also have determinant $\sqrt{|\Delta|}$. 

\begin{algorithm}[ht]
\caption{Classical oracle evaluation (without phase)}
\begin{algorithmic}[1]\label{alg:oracle}
\REQUIRE $(u_2,\cdots,u_{n_1+n_2},v_1,\cdots,v_{|S|})$.
\ENSURE The $E$-ideal corresponding to $\phi(u_1,\cdots,u'_{n_1+n_2})\cdot\OOO\prod_j\pg_j^{-v_j}$.
\STATE Compute $u_1$ according to~\eqref{eq:last-coordinate}
\STATE Compute $A_{-1}$ using the formula $e^{x}\simeq \sum \frac{x^i}{i!}$.
\STATE Compute the $A_j$ using~\eqref{eq:def_Aj}.
\STATE Compute the $B_{j,k}$ using~\eqref{eq:def_Bj}.
\STATE For each $k\leq \log_2(v)$, $B_k\leftarrow \prod_j B_{j,k}$.
\RETURN $A_{-1}\cdot\prod_j A_j^{2^j}\cdot \prod_k B_k^{2^k}$. 
\end{algorithmic}
\end{algorithm}

\begin{proposition}
Algorithm~\ref{alg:oracle} is correct and involves a polynomial number of multiplications between 
$E$-ideals of determinant $\sqrt{|\Delta|}$. 
\end{proposition}

\subsection{$E$-ideal arithmetic}\label{subsec:arithmetic}

The arithmetic between $E$-ideals is directly inspired from the arithmetic 
between ideals in a number field. To evaluate our classical oracle, we need 
an efficient implementation of the $E$-ideal multiplication. Let 
$A = \oplus_{j\leq n} \Z a_j$ and 
$B = \oplus_{k\leq n} \Z b_k$ be $E$-ideals generated by the $a_j,b_k\in E$. Then 
the $E$-ideal $A\cdot B$ is the lattice generated by the $n^2$ elements 
$(a_j\cdot b_k)_{j,k\leq n}$. The multiplication of two $E$-ideals 
can be described by the two following steps:
\begin{enumerate}
\item Calculate all the cross terms $a_j\cdot b_k$ for $j,k\leq n$. 
\item Compute a basis $(c_j)_{j\leq n}$ of $\sum_{j,k}\Z a_j\cdot b_k$. 
\end{enumerate}
The main challenge of $E$-ideal multiplication is that we need to deal 
with rational approximations of lattices. We need to estimate how much 
precision is needed to ensure accuracy, and how much precision is lost 
after each operation. We employ the same strategy as in~\cite{STOC2014}, which uses fixed point rational  approximations such that  $\left| \frac{\widehat{a}}{2^{q_0}}-a\right|\leq \frac{1}{2^{q}}$ where the precision $q$ deteriorates throughout the computation. 

\paragraph{Multiplication of approximate elements}

$E$-ideals can be seen as lattices in $\R^m$ where $m = n_1 + 4n_2$ by 
mapping elements in $E$ of the form $(u_1,\cdots,u_{n_1,n_2})$ to 
\begin{equation}\label{eq:map_Rm}
(u_1,\cdots,n_{n_1},\Re(u_{n_1+1}),\Im(u_{n_1+1}) , \Re(u_{n_1+1}),-\Im(u_{n_1+1})
,\cdots,)\R^{m}.
\end{equation}
We approximate each coordinate $a\in\R$ of such vector by $\widehat{a}/2^{q_0}$ with 
$\widehat{a}\in\Z$ and $q_0\in\Z_{>0}$ satisfying 
$\left| \frac{\widehat{a}}{2^{q_0}} - a\right| \leq \frac{1}{2^q}$ for some $q\leq q_0+1$. Then the multiplication of 
$q$-bit approximations of $a,b\in\R$ with $\log_2(|a|),\log_2(|b|)\leq c$ satisfies 
$$\left| \frac{\widehat{a}\widehat{b}}{2^{2q_0}} - ab\right| 
=\left| \left( \frac{\widehat{a}}{2^{q_0}}-a\right)\frac{\widehat{b}}{2^{q_0}} 
+ \left( \frac{\widehat{b}}{2^{q_0}}-b\right)a\right| \leq 
\frac{2^c}{2^q} + \frac{2^{c}}{2^q} = \frac{1}{2^{q-c-1}}.$$
This gives us a rational number approximating $ab$, but the denominator 
increases, which reduces the efficiency of the computation of a 
basis of the $E$-ideal generated by the $n^2$ products. Let 
$\widehat{ab} := \left\lfloor 2^{q_0}\left(\frac{\widehat{a}\widehat{b}}{2^{2q_0}}\right)\right\rceil$, 
then 
$$\left| \frac{\widehat{ab}}{2^{q_0}} - ab\right| 
\leq \left| \frac{\widehat{ab}}{2^{q_0}} - \frac{\widehat{a}\widehat{b}}{2^{2q_0}}\right|
+ \left|\frac{\widehat{a}\widehat{b}}{2^{2q_0}} - ab\right| 
\leq \frac{1}{2^{q_0+1}} + \frac{1}{2^{q-c-1}} \leq \frac{1}{2^{q-c-1}} + \frac{1}{2^{q-c-1}} = \frac{1}{2^{q-c-2}}.$$
This gives us the approximation of the multiplication of two real numbers. Although 
we approximate vectors in $E$ by vectors in $\R^m$, the pointwise multiplication of 
the $m-n_1$ last coordinates correspond to multiplications between complex numbers. 
The addition involved in the complex number multiplication 
$(a + ib)(c + id) = (ac-bd) + i(ad + bc)$ induces the loss of an extra 
bit of precision. If $a,b\in\R$ are approximated by $\widehat{a},\widehat{b}$ with 
precision $q$, then $\widehat{a+b}:= (\widehat{a}+\widehat{b})$ 
satisfies $\left| \frac{\widehat{a+b}}{2^{q_0}} - (a+b)\right| \leq \frac{1}{2^{q-1}}$. We summarize pointwise multiplication of approximations 
of elements in Algorithm

\begin{algorithm}[ht]
\caption{Multiplication between approximations of elements in $E$}
\begin{algorithmic}[1]\label{alg:mult_E}
\REQUIRE $\widehat{a},\widehat{b}\in\Z^{m}$ such that 
$\left| \frac{\widehat{a}_j}{2^{q_0}} - a_j\right|\leq \frac{1}{2^q}$ and 
$\left| \frac{\widehat{b}_j}{2^{q_0}} - b_j\right|\leq \frac{1}{2^q}$ where 
$(a_j),(b_j)$ are the entries of $a,b\in E$ and $c\geq \log_2(\|a\|),\log_2(\|a\|)$.  
\ENSURE An approximation $\widehat{ab}\in\Z^m$ of the vector $ab\in \R^m$ 
with precision $q':= q - c - 3$. 
\STATE \textbf{For} $j\leq n_1$ \textbf{do}  
$\widehat{ab}_j := \left\lfloor 2^{q_0}\left(\frac{\widehat{a}_j\widehat{b}_j}{2^{2q_0}}\right)\right\rceil$.
\WHILE{$n_1<j\leq m$}
\STATE $w := \left\lfloor 2^{q_0}\left(\frac{\widehat{a}_j\widehat{b}_{j}}{2^{2q_0}}\right)\right\rceil$ , 
$x := \left\lfloor 2^{q_0}\left(\frac{\widehat{a}_{j+1}\widehat{b}_{j+1}}{2^{2q_0}}\right)\right\rceil$.
\STATE $y := \left\lfloor 2^{q_0}\left(\frac{\widehat{a}_j\widehat{b}_{j+1}}{2^{2q_0}}\right)\right\rceil$
$z := \left\lfloor 2^{q_0}\left(\frac{\widehat{a}_{j+1}\widehat{b}_{j}}{2^{2q_0}}\right)\right\rceil$.
\STATE $\widehat{ab}_j := w-x$, $\widehat{ab}_{j+1} := y+z$, $j\leftarrow j+2$. 
\ENDWHILE
\RETURN $(\widehat{ab}_j)$. 
\end{algorithmic}
\end{algorithm}

\begin{lemma}
    Algorithm~\ref{alg:mult_E} returns a vector $\widehat{ab}\in\Z^m$ whose entries approximates those of $ab\in E$ with precision $q-c-3$ in time $\tilde{O}(m(q_0 + c))$. Additionally, this vector satisfies 
    $$\left\| \frac{\widehat{ab}}{2^{q_0}} - ab\right\| \leq \frac{\sqrt{m}}{2^{q-c-3}}$$
\end{lemma}


\paragraph{Computing a basis from a generating set}

Let $\Lambda$ be an $E$-ideal for which we want to find a basis of short vectors in polynomial time. 
As the Euclidean norm is preserved by the mapping of $\Lambda$ in 
$\R^m$, 
this problem boils down to computing a short basis of 
an ideal in $\R^m$. Since the original description of the LLL reduction algorithm~\cite{LLL}, the 
problem of finding a short basis (up to an approximation factor) of a lattice in polynomial time 
is well understood. The difficulty in this context is that we are dealing with rational approximations 
of real numbers. Let $a = (a_1,\cdots,a_m)\in\R^m$ and $q_0+1\geq q\geq 0$, we say that 
$\widehat{a} = (\widehat{a}_1,\cdots,\widehat{a}_m)\in\Z^m$ is an approximation of $a$ with precision $q$ if 
$\forall j\leq m, \ \left| \frac{\widehat{a}_j}{2^{q_0}} - a_j\right| \leq \frac{1}{2^q}$. Given an approximate generating set for 
the lattice $\Lambda\in\R^m$,  
we want to compute a basis of short vectors that approximates a basis of short vectors for $\Lambda\in\R^m$. 
We rely on a result from Buchmann and Kessler~\cite{BK93} and its modification by Eisentr\"{a}ger, Halgren, 
Kitaev and Song~\cite{STOC2014}. 

Let $\widehat{a}_1,\cdots,\widehat{a}_k\in\Z^m$ be rational approximations of $a_1,\cdots,a_k\in\R^m$ 
of precision $q$ (and denominator $q_0\geq q-1$). Let $r\leq k$ be the rank of the lattice generated by 
$(a_j)_{j\leq k}$. The approach 
described in~\cite{BK93} consists of applying the LLL reduction algorithm to the rank $k$ lattice 
generated by the independent vectors $\tilde{a}_j := (e_j,\widehat{a}_j), j\leq k$ where 
$e_j$ is the $j$-th unit vector of $\Z^k$. The LLL algorithm outputs vectors 
$\tilde{b}_j = (m_j,\widehat{b}_j), j\leq k$ such that if the input precision $q$ is large 
enough, $m_1,\cdots,m_{k-r}$ are independent relations for $a_1,\cdots,a_k$ (i.e. 
$\sum_l m_l^{(j)}a_j^{(l)} = 0$) and the vectors 
$b_j = \sum_j m_{k-r+j}^{(l)} a_l, j\leq r$ are a basis for the lattice $\sum_j \Z a_j$. 

The following proposition states our modification of the result of~\cite{BK93} incorporating the cases where $q_0\geq q$.

\begin{algorithm}[ht]
\caption{Computation of a basis from an approximate generating set}
\begin{algorithmic}[1]\label{alg:basis}
\REQUIRE Approximate vectors $(\widehat{a}_j)_{j\leq k}\in\Z^m$ of $(a_j)_{j\leq k}\in\R^m$ of precision 
$q$ and denominator $q_0$. 
\ENSURE Approximate vectors $(\widehat{b}_j)_{j\leq r}\in\Z^m$ of a basis of $\sum_j\Z a_j$. 
\STATE \textbf{for } $j\leq k$ \textbf{do} $\tilde{a}_j\leftarrow (e_j,\widehat{a}_j)$. 
\STATE Apply LLL to the $\tilde{a}_j$ and get $\tilde{b'}_j:= (m'_j,\widehat{b'}_j)$.
\STATE Apply LLL to the $\widehat{b'}_j$ for $k-r+1\leq j\leq k$. and get $\widehat{b}_{j-k+r}$ for $k-r+1\leq j\leq k$.
\RETURN $\widehat{b}_{l}$ for $l\leq r$. 
\end{algorithmic}
\end{algorithm}

\begin{proposition}[Theorem~C.5 of~\cite{EHKS14long}]\label{prop:cost-mult}
Let $a_1\ldots,a_k\in \R^m$ be a generating set for a lattice $L$ of rank $r$. Let $\widehat{a_1},\ldots,\widehat{a_k}\in\Z^m$ be rational approximations of the $a_i$ with precision $q$ and denominator $q_0$ (i.e. $\forall i\leq k,\ \|\widehat{a_i}/2^{q_0} - a_i\|\leq \sqrt{m}/2^{2^q}$). Assume $\mu$ is a lower bound on $\lambda_1(L)$, and that 
$$2^q\geq \left( k2^{\frac{k+1}{2}}\max\|a_i\|\right)^r / \left(\mu\det(L)^2\right).$$
Then Algorithm~\ref{alg:basis} returns approximations $\widehat{c_1},\ldots,\widehat{c_r}\in\Z^r$ of a basis $c_1,\ldots,c_r$ of $L$ that satisfies 
$$\forall i\leq r,\ \|c_i\|\leq \left( \sqrt{mk}+2\right)2^{\frac{k-1}{2}}\lambda_i(L).$$
Additionally, the precision of the output satisfies 
$\|\widehat{c_i}/2^{q_0} - c_i\|\leq rk\gamma_1\gamma_3\sqrt{m}/2^q$ where 
\begin{itemize}
    \item $\gamma_1\leq k^22^{\frac{k+1}{2}}\frac{\max_i\|a_i\|}{\det(L)}$
    \item $\gamma_3 \leq \frac{4k\left( k2^{k/2}\max_i\|a_i\|\right)^r}{\det(L)^2}$.
\end{itemize}
The cost of this procedure is $$\operatorname{Cost}_{\text{LLL}}(k , k, \log_2(\max_i\|a_i\|) + q_0) + \operatorname{Cost}_{\text{LLL}}\left(r ,k, \log_2\left( k(\sqrt{km}+2)2^{\frac{k-1}{2}}\max_i\|a_i\|\right) + q_0\right),$$
where 
$\operatorname{Cost}_{\text{LLL}}(r,d,b)$ denotes the cost of running the LLL algorithm on $r$ linearly independent vectors of $\Z^d$ with entries of bit size bounded by $b$.
\end{proposition}

\begin{proof}
    The bounds on the size of the entries and on the precision of the output are exactly the statement made in~\cite[Th. C.5]{EHKS14long}. To derive the bound on the cost of the procedure, one needs to use information in the proof of~\cite[Th. C.5]{EHKS14long} regarding the size of the entries of the vectors given as input to the second call to LLL. Indeed, it is proved that the vectors have length bounded by 
    $k(\sqrt{km}+2)2^{\frac{k-1}{2}}\max_i\|a_i\|$.
\end{proof}

\subsection{Complexity of the classical computation}\label{subsec:complexity}

To estimate the asymptotic complexity of the classical oracle, we need to combine the results of 
Section~\ref{subsec:comp_split} and Section~\ref{subsec:arithmetic}. Let 
$(y,v_1,\cdots,v_|S|)\in \R^ {n_1+n_2-1}\times \Z_2^{n_1}\times (\R/\Z)^{n_2}\times \Z^{|S|}$. We want to compute a poly-size basis of
$$f_c(y,v_1,\cdots,v_{|S|}) = \phi(y)\cdot \OOO \cdot \pg_1^{-v_1}\cdots \pg_{|S|}^{-v_{|S|}},$$
in polynomial time in $\max_j\{\log(|y_j|)\}$, $|S|$, $\max_j\{\log(p_j)\}$, 
$\max_j\{\log|v_j|\}$, $m$, and 
$\log|\Delta|$. 

\begin{theorem}\label{th:classical-oracle_cost}
Let $K$ be a number field of degree $n$ and discriminant $\Delta$. Let $S$ be a set of prime ideals of $K$. The gate complexity of the classical oracle is in 
$$
 O\left( \left(\log_2(v)(|S|+\log_2(v)) + \log_2(r)^2\right)C_{\text{gates}}(n^2,\beta\right),
$$
and the qubit requirement is in 
$$O\left(\max(|S|,\log_2(v),\log_2(r))C_{\text{qubits}}(n^2,\beta\right)$$
%
where 
$v = \max_i|v_i|$, $r = \max_i|y_i|$, 
$$\beta \in \tilde{O} \left( n(n^2+\log|\Delta|)\cdot\max(\log|S|+\log(v),\log(r))\right),$$
is a bound on the bit size of the integer vectors given to the LLL algorithm, and the costs $C_{\text{gates}}(k,b),C_{\text{qubits}}(k,b)$ denote the gate cost (resp. memory cost) of the LLL algorithm on input an $k$-rank lattice in $\Z^k$ with entries of bit size bounded by $b$. As usual $\tilde{O}$ denotes the complexity where the logarithmic factors are omitted.
\end{theorem}


\begin{proof}
Given how we split 
up the computation, we only multiply $E$-ideals of determinant $\sqrt{|\Delta|}$. This means that once the original precision $q_0$ is chosen, the cost of each ideal multiplication is bounded. Let $C_{\text{gates}}$ be a bound on the number of quantum gates required for the multiplication of two ideals, and $C_{\text{qubits}}$ be a bound on the amount of memory required to multiply two ideals. On input $(u,v)\in\R^{n_1+n_2}\times \Z^{|S|}$ the evaluation of $f_c$ consists in the calculation of $\prod_{k\leq \log_2(r)}A_k^{2^k}$ and $\prod_{k\leq \log_2(v)}B_k^{2^k}$ where $\log_2(r) \leq \max_j(\log_2(|u_i|))$, $\log_2(v) \leq \max_j(v_i)$, and the $A_k,B_k$ are defined in Section~\ref{subsec:comp_split}.

We first analyze the cost of computing $\prod_{k\leq \log_2(v)}B_k^{2^k}$. For each $k\leq \log_2(v)$, we compute $B_k := \prod_{j\leq |S|}B^{b_j^{(k)}}_{j,k}$. To minimize the length of the multiplication chain, we use a multiplication tree of depth $\log_2(|S|)$. This costs $O(|S|C_{\text{gates}})$ gates and uses $O(|S|C_{\text{qubits}})$ qubits. Then we compute $B_{k}^{2^k}$ by performing $k$ squarings with $O(kC_{\text{gates}})$ gates and $O(kC_{\text{qubits}})$ qubits. This procedure is repeated for all $k\leq \log_2(v)$ using a total of 
$O(\log_2(v)(|S| + \log_2(v))C_{\text{gates}})$ gates and $O(\max(|S|,\log_2(v))C_{\text{qubits}})$ qubits. Finally, we compute the product $\prod_{k\leq \log_2(v)}B_k^{2^k}$ via a product tree of depth $\log_2\log_2(v)$ using $O(\log_2(v)C_{\text{gates}})$ gates and $O(\log_2(v)C_{\text{qubits}})$ qubits. In summary, $\prod_{k\leq \log_2(v)}B_k^{2^k}$ is computed using $O(\log_2(v)(|S| + \log_2(v))C_{\text{gates}})$ gates and $O(\max(|S|,\log_2(v))C_{\text{qubits}})$ qubits. The longest multiplication chain leading to this value is of length $\log_2(|S|) + \log_2(v) + \log_2\log_2(v)$. 

Next, we analyze the cost of computing $\prod_{k\leq \log_2(r)}A_k^{2^k}$. We follow the same strategy as above. First we compute all the $A_{k}^{2^{k}}$ for $k\leq \log_2(r)$ using $O(\log_2(r)^2C_{\text{gates}})$ gates and $O(\log_2(r)C_{\text{qubits}})$ qubits. The product  $\prod_{k\leq \log_2(r)}A_k^{2^k}$ is then computed with a product tree of depth $\log_2\log_2(r)$ using an extra $O(\log_2(r)C_{\text{gates}})$ gates. The total number of gates required is in $O(\log_2(r)^2C_{\text{gates}})$, and the number of required qubits in $O(\log_2(r)C_{\text{qubits}})$. The longest multiplication chain in of length $\log_2(r) + \log_2\log_2(r)$.

In summary, the gate cost, memory cost and longest multiplication chains of the computation satisfy:
\begin{itemize}
    \item Gate cost in $O\left( \left(\log_2(v)(|S|+\log_2(v)) + \log_2(r)^2\right)C_{\text{gates}}\right)$.
    \item Memory cost in $O\left(\max(|S|,\log_2(v),\log_2(r))C_{\text{qubits}}\right)$.
    \item Longest multiplication chain: $\max(\log_2|S| + \log_2(v) + \log_2\log_2(v),\log_2(r) + \log_2\log_2(r))$.
\end{itemize}
To determine the cost of each multiplication, we need to evaluate the size of the entries of the matrices that are given as input to the LLL algorithm. This depends on the precision we require. For Proposition~\ref{prop:cost-mult} to apply, we need that the final precision $q$ satisfy $2^q\geq \left( k2^{\frac{k+1}{2}}\max\|a_i\|\right)^r / \left(\mu\det(L)^2\right)$ where $\mu$ is a lower bound on $\lambda_1(L)$. We have that $\det(L) = \sqrt{|\Delta|}$ and $\lambda_1(L)\geq \sqrt{n}$. Moreover, we can always assume that matrix given as input is reduced, which means that $\|a_i\|\leq \left( \sqrt{nk}+2\right)2^{\frac{k-1}{2}}\lambda_i(L)\leq  \left( \sqrt{nk}+2\right)2^{\frac{k-1}{2}}\sqrt{n|\Delta|}$. At each ideal multiplication, the number $q_{\text{loss}}$ of bits of precision we lose satisfies 
$$
q_{\text{loss}}\leq \log_2\left( \left( \sqrt{nk}+2\right)2^{\frac{k-1}{2}}\sqrt{n|\Delta|}\right)  + \log_2\left( n^{3.5}\gamma_1\gamma_3\right),
$$
where $\gamma_1,\gamma_3$ are defined in Proposition~\ref{prop:cost-mult}. If we define $\ell_{\text{mult}} := \max(\log_2|S| + \log_2(v) + \log_2\log_2(v),\log_2(r) + \log_2\log_2(r))$ to be the length of the longest multiplication chain, and the target final precision 
$q_{\text{final}}:= n\log_2\left( k\sqrt{n}(\sqrt{nk}+2)2^k\sqrt{|\Delta|}\right)$, then the initial precision is $q_0 := q_{\text{final}} + \ell_{\text{mult}}q_{\text{loss}}$. To simplify the asymptotic estimate of the cost of the multiplication of ideals, we notice that the bound $\alpha$ on the bit size of the vectors of the basis of $L$ given as input to the first LLL solver, and the bound $\alpha'$ on the bit size of the basis vectors given to the second LLL solver both satisfy:
$$\alpha,\alpha'\in \tilde{O}\left( n^2 + \log_2|\Delta|\right).$$
This means that the cost of each multiplication is dominated by the cost of LLL on input a rank and dimension $n^2$ lattice with vectors of bit size in $\tilde{O}\left( n^2 + \log_2|\Delta| + q_0\right)$. We then notice that 
$$q_0\in \tilde{O} \left( n(n^2+\log|\Delta|)+\max(\log|S|+\log(v),\log(r))\right),$$
which asymptotically dominates the bit size of the integer vectors given as input to the LLL algorithm.

\end{proof}

\subsection{The quantum encoding of $e^{\textbf{t}}\underline{\OO}\prod_{\pg\in S}\pg^{-v_i}$}\label{sec:quantum-encoding}

\noindent Let $g_s(\cdot)$ be the Gaussian function $g_s(x): = e^{-\pi \|x\|^2/s^2}, x\in \bR^n$. 
For any set $S\subset \bR^n$, denote $g_s(S): =\sum_{x\in S} g_s(x)$. 
Given a lattice $L$, the quantum encoding maps $L$ to the lattice Gaussian state via
$$\begin{CD}
\{\text{Lattices over }E\} \quad @>{f_q}>> \cS  \text{ (unit vectors in a Hilbert space)}\\
L @>>> \ket{L}: = \gamma \sum_{v\in L} g_s(v) \ket{\strn{v}}
\end{CD},$$
where $\gamma$ is a factor that normalized the state. Here $\ket{\strn{v}}$ is the 
straddle encoding of a real-valued vector $v\in\bR^n$, as defined in~\cite{STOC2014}.  Intuitively, we 
discretize the space $\bR^n$ by a grid $\nu\bZ^n$, and we encode the information about 
$v$ by a superposition over all grid nodes surrounding $v$. Specifically, for the one-dimensional case, 
the straddle encoding of a real number is 
$$ x \in \bR \mapsto \ket{\str{x}}: = \cos(\frac \pi 2 t) \ket{k} + \sin(\frac\pi 2 t) \ket{k+1} \, ,$$ 
where $k: = \lfloor x/\nu \rfloor$ denotes the nearest grid point no bigger than 
$x$ and $t: = x/\nu - k$ denotes the (scaled) offset. Repeat this for each coordinate of $v=(v_1,\ldots, v_n)$ 
we get $\ket{\strn{v}} := \bigotimes_{i=1}^n \ket{\str{v_i}}$. We recall some properties about 
straddle encoding from~\cite{STOC2014}. This will be useful to prove 
the HSP properties of our function. 

\begin{fact}
Let $v,w\in \bR^n$. The following hold
\begin{itemize}
	\item $\| \ket{\strn{v}} - \ket{\strn{w}} \| \leq \frac \pi {2\nu} \sqrt n \cdot \| v - w\|$. 
	\item If $\| v - w \| \geq 2\sqrt n \nu$, then $\inp{\strn{v}}{\strn{w}} = 0$.
\end{itemize}
\end{fact}

\noindent In our lattice Gaussian states, we will always make 
sure $\lambda_1(L)> 2\sqrt n \nu$ so that 
$$\inp{\strn{v}}{\strn{u}} = 0 \text{ whenever } v \neq u \, .$$ In this case 
we can compute the normalization factor $\gamma = \left( g_{\frac{s}{\sqrt 2}}(L)\right)^{-1/2}$. 
As shown in~\cite{STOC2014}, one can efficiently compute $f_q$ if the lattice satisfies certain 
conditions and a good basis is given (e.g., $L$ is LLL-reduced). 
Namely there is an efficient quantum circuit creating lattice Gaussian states. 
We state this result with gate and qubit count estimation below and will invoke it as a black-box. 

\begin{fact}
Let $L$ be an LLL-reduced basis. Assume that 
$\lambda_1(L)\geq \lambda_0$, $\det(L) \leq d_0 $ and $s\geq n^{n/2+1} 2^{3n} \lambda_0^{-n+1} d_0$. Let $\ket{L} = \gamma \sum_{v\in L} g_s(v) \ket{\strn{v}}$. 
There is a quantum circuit that takes $L$ as input and produces a state $\ket{\psi}$ 
such that $\| \ket{\psi} - \ket{L} \| \le 2^{-n}$. The quantum circuit has gate count $O(n^3 + n\log (1/\nu))$ and qubit count $O(n \log (\sqrt n \|L^{-1}\| s) ) = O(n^2 (\log n + \log (1/\lambda_0))$. 
\end{fact}

\section{Pseudoinjectivity of
  $(\textbf{y},\textbf{v})\mapsto
  |\phi(\textbf{y}) \underline{\OO}\prod_{\pg\in
    S}\pg^{-v_i}\rangle$}
\label{ssec:pseudoinjectivity}

\begin{theorem}\label{thm:pseudoinjectivity}
    Let $f$ be the function $ G = \R^{n_1 + n_2 - 1}\times \Z_2^{n_1} \times \left(\R/\Z\right)^{n_2}\times \Z^s \rightarrow \mathcal{H}$ defined by $(\textbf{y},\textbf{v})\mapsto
  |\phi(\textbf{y}) \underline{\OO}\prod_{\pg\in
    S}\pg^{-v_i}\rangle$. There is $r,\varepsilon>0$ such that 
    $$
    d_{G/L}(x,y) := \min_{v\in U^G_S}\|x-y-v\|\ge r \Rightarrow \bigl|\langle{f(x)}|{f(y)}\rangle\bigr|\le\varepsilon
    $$
\end{theorem}

Our proof relies on some statements on lattices available in~\cite{EHKS14long}. 
As in~\cite[Sec. E.2]{EHKS14long}, we first introduce a central notion called the \emph{approximate intersecting sublattice} of two lattices $L$ and $L'$ in $\R^m$. 

\begin{Definition}[$\delta$-approximate intersecting sublattice]
Let $L$ and $L'$ be two lattices of dimension $n$ in $\bR^m$. Let 
$Y:= \{ (x,x'): x\in L_R, x' \in L'_R, \|x - x'\|\leq \delta\}$ and $X: = Y|_1$ ($X': = Y|_2$) 
be the corresponding set of points $x$ (resp. $x'$). Define $\aisub:=\langle X\rangle$ ($\aisub':=\langle X' \rangle$ resp.) 
be the sublattice generated by points in $X$ ($X'$ resp.). We call 
$\aisub$ ($\aisub'$) the $\delta$-approximate intersecting sublattice of $L$ (resp. $L'$) 
between $L$ and $L'$. 
\label{def:aisub}
\end{Definition}

\noindent Here $L_R = L \cap \mathbf{B}_R$ are the lattice points inside a sphere of radius $R=\sqrt n s$, where $s$ is the Gaussian width in the lattice Gaussian state. This definition indeed captures the overlap (up to $\delta$-approximation) between two lattices. 
Intuitively, $\aisub$ and $\aisub'$ can be paired up that are ``close'', and all the other pairs of points 
will be ``far'' apart. This overlap is the main contribution to the inner product between the 
quantum encoding of two lattices, and we show that if it generates a proper sublattice, 
we can bound the scalar product. This is formalized below as shown in~\cite{STOC2014}. We sketch a proof for completeness. 

\begin{fact}[Lemma E.6 of~\cite{EHKS14long}]\label{fact:strprop}
Let $L,L',\aisub$ and $\aisub'$ be as in Definition~\ref{def:aisub}. Suppose that: 
$\lambda_1\geq \lambda$, $\lambda_1' \geq \lambda$. Then there is a one-one 
correspondence $h:\aisub \to \aisub'$ such that 
\begin{itemize}
	\item $\forall x \in \aisub, \| x - h(x)\| \leq \beta \|x\|$ with 
	$\beta:= n(\sqrt n R/\lambda)^n \cdot \frac \delta R$;
	\item For any $x \in L_R $ and any $ x' \in L'_R$, if $x' \neq h(x)$ (in particular if $x\notin \aisub$ or $x'\notin \aisub'$,  $\|x - x'\| > \delta$. 
\end{itemize}
\end{fact}

\begin{proof} (Sketch) Pick $x_i \in X: i = 1,\ldots, n$ that are linearly independent and 
let $x_i'$ be the corresponding points in $X'$. Let $h: x_i \mapsto x_i'$ and this extends 
to a linear map from $\bar\aisub$ to $\bar\aisub'$. The second property holds immediately 
by definition. To show the first one, let $x\in \aisub$ and write it as 
$x= \sum_i \alpha_i x_i, \alpha_i \in \bR$. Using Cramer's rule, Hadamard inequality 
and Minkowski's second theorem, one can get $|\alpha_i| \leq (\sqrt n R /\lambda)^n \frac{\vnorm{x}}{R}$. 
Therefore $\vnorm{x - h(x)} = \vnorm {\sum_i \alpha_i (x_i - h(x_i))} \leq \sum_i |\alpha_i| \delta \leq \beta \vnorm{x}$ 
with $\beta=n(\sqrt n R/\lambda)^n\cdot \frac \delta R$. 
 
\end{proof} 

\noindent If we pick the straddle encoding fine enough such that $2\sqrt n \nu < \delta$, 
it follows that the inner product between their quantum encodings will be solely contributed by $\aisub$ and $\aisub'$. 
In particular:

\begin{fact}[Lemma E.7 of~\cite{EHKS14long}]\label{fac:epsilon-condition}
Let $\mathcal{L}$ and $\mathcal{L}'$ be two $E$-ideals with 
$\max\{\det(\mathcal{L}), \det(\mathcal{L}')\} \leq d$ and 
$\min{\lambda_1(\mathcal{L}), \lambda_1(\mathcal{L}')}\ge \lambda$. Let 
$\aisub$ and $\aisub'$ be the $\delta$-intersecting sublattices of $\cL$ and $\cL'$ respectively, as defined in Definition~\ref{def:aisub}. 
If $\aisub \subsetneq \mathcal{L}$ (which implies $\aisub' \subsetneq \mathcal{L}'$), 
then $\inp{\mathcal{L}}{\mathcal{L}'} \leq 3/4$ whenever $s\geq 4\pi n^{n/2+3}  d /\lambda^{n-1}$. 
\label{lemma:inppsub}
\end{fact}




\noindent The two previous claims give us a sufficient condition for 
$\inp{\mathcal{L}}{\mathcal{L}'}\leq 3/4$. To prove the $(r,\varepsilon)$-condition, 
we need to relate the properties of $\Lambda$ to our notion of distance between the preimages 
in $G$. We first prove a sufficient condition on $\dist(\cL,\cL')$ in Lemma~\ref{lemma:psub}, 
which ensures that the approximate intersecting sublattices $\aisub$ and $\aisub'$ be \emph{proper}. 

\begin{lemma}\label{lem:1}
If $\dist(\mathcal{L},\mathcal{L}') \geq r= \frac{1}{2\sqrt n |\Delta|}$ and $\beta < \frac{1}{20n^{n+2}|\Delta|}$, then 
 the $\delta$-intersecting sublattices $\aisub$ and $\aisub'$ of $\cL$ and $\cL'$ respectively, as defined in Definition~\ref{def:aisub}, become proper sublattices. Namely 
$\aisub\subsetneq \mathcal{L}$ and $\aisub' \subsetneq \mathcal{L}'$.

On the other hand, if $\aisub= \mathcal{L}$ and $\aisub'= \mathcal{L}'$, then 
there is $W$ satisfying $M_{\cL'}=M_\cL W$ for any bases $M_{\cL'},M_\cL$ of $\cL',\cL$ that is of the 
form $W = e^{\diag(a)}$ for some $a$ with $\|a\|\leq \frac{1}{4\sqrt{n}|\Delta|}$.

\label{lemma:psub} 
\end{lemma}

\begin{proof}
Suppose for contradiction that $\aisub = \mathcal{L}$ and $\aisub' = \mathcal{L}'$. Let 
$M_h$ be the matrix induced by $h$ (wrt to some choice of basis for $\cL$ and $\cL'$). First we claim that 
$\onorm{M_h - I} \leq \beta^{(1)} : = n^{n+1} \beta$. To show this, we pick a short basis $(v_1, \ldots, v_n)$ for 
$\mathcal{L}$ such that $\|v_k\| \leq \sqrt k \lambda_k(\mathcal{L})$ for $k\leq n$, which 
always exists. 
Then any $w \in \bR^n$ with $\vnorm{w} = 1$ can be written as $w = \sum_i \alpha_i v_i, \alpha_i \in \bR$. 
By Cramer's rule we have 
$$|\alpha_i| = \left|\frac{\det(v_1, \ldots, v_{i-1}, w, v_{i+1}, \ldots, v_n)}{\det(v_1,\ldots, v_n)} \right| 
\leq \frac{(\sqrt n)^n \Pi_{j\neq i} \lambda_j(\mathcal{L})}{\sqrt i \det(\mathcal{L})} \leq n^{n}/{\sqrt i \lambda_i(\mathcal{L})}\, .$$ 
The first inequality uses Hadamard's inequality and the second inequality invokes Minkowski's second theorem 
$\Pi_j \lambda_j(\mathcal{L}) \leq n^{n/2}\det(\mathcal{L})$. Then 
$$\vnorm{w(M_h - I)} = \left\|{\sum_i \alpha_i (h(v_i) - v_i)}\right\| \leq \sum_i |\alpha_i|\cdot \vnorm{h(v_i) - v_i} 
\leq n\cdot \frac{n^n}{\sqrt i \lambda_i(\mathcal{L})}  \cdot \beta \|v_i\|\leq n^{n+1}\beta \, .$$ This 
implies that $\onorm{M_h - I} \leq \beta^{(1)}$. 

Next, by choosing $W:= M_h$, we have 
$\fnorm{W - I} \leq \beta^{(2)}:=\sqrt{n}\beta^{(1)} = n^{n+3/2}\beta$, and $M_{\cL'} = M_{\cL}W$ where $M_{\cL'}$ (resp. $M_{\cL}$) are 
matrices for the choice of basis of $\mathcal{L}'$ (resp. $\mathcal{L}$) that corresponds to $M_h$ (i.e. $M_{\cL'} = M_{\cL}M_h$). 

Then, since $\beta^{(2)} < \left( 20\sqrt{n}|\Delta|\right)^{-1}$, $W$ is necessarily diagonal (see Claim~\ref{claim:diag_matrix}), and 
hence $M_{\cL'} = M_{\cL}e^{\diag(a_i)} $ with 
$\fnorm{e^{\diag(a_i)} - I}\leq \beta^{(2)}$. This 
implies that\footnote{Let $A=\diag{a_i}$. Observe that $\fnorm{A}\leq 1$ and in 
this case $\sum_{k=2}^\infty \frac{\fnorm{A^k}}{k!}\leq \fnorm{A}\cdot \sum_{k=2}^\infty \frac{1}{k!} = (e-2)\fnorm{A}$. 
Hence $\fnorm{e^A - I} \geq \fnorm{A} - \sum_{k=2}^\infty \frac{\fnorm{A^k}}{k!}\geq 0.2 \fnorm{A}$.} 
$$ \vnorm{a}_E =\|\diag(a_i)\|_2  \leq 5\beta^{(2)} < \frac{1}{4\sqrt n |\Delta|}$$ 
when $\beta < \frac{1}{2n^{n+2}|\Delta|}$, and hence since $\|\diag(a_i)\|_2\geq \dist(\cL,\cL')$, it contradicts the hypothesis that 
$\dist(\mathcal{L},\mathcal{L}') \ge r$. 

%
%
\end{proof}

The following claim is taken from an unpublished version of~\cite{EHKS14long}.
\begin{claim}[Sections E.2 and E.3 of~\cite{EHKS14long}]
Let $\cL$ (resp. $\cL'$) be $E$-ideals of norm 1 admitting a basis represented by the matrix $M_\cL$ (resp. $M_{\cL'}$) 
satisfying $M_{\cL'} = M_\cL W$ for some matrix $W$. If $\|W-I\| < \left( 2\sqrt{n}|\Delta|\right)^{-1}$, then 
$W = \diag(z)$ for some $z\in E$. 
\label{claim:diag_matrix}
\end{claim}

\begin{proof}
For completeness, we reproduce the proof of this statement as it is presented in~\cite{EHKS14long}. 
The matrix $W$ is of the form $\diag(z)$ if and only if it commutes with all matrices of the 
form $\diag(z)$. To check that, it suffices to show that $M_\cL(W\diag(\omega_j) - \diag(\omega_j) W) = 0$ 
where $(\omega_k)_{k\leq n}$ is an integral basis of $\OO$. Indeed both the $\omega_k$ and 
the rows $b_1,\cdots,b_n$ of $M_\cL$ are linearly independent. We can assume that 
$\|\omega_j\|_E\leq \lambda_n(\OO)$ and $\|b_j\|_E\leq \lambda_n(\cL)$. Moreover, we know that 
$$\lambda_1(\cL')\geq \sqrt{n}\Nm(\cL')^{1/n},\ \ \lambda_n(\OO)\leq \sqrt{n|\Delta|},\ \ 
\lambda_n(\cL)\leq \sqrt{n|\Delta|}\Nm(\cL)^{1/n}$$.
Therefore, since $\Nm(\cL),\Nm(\cL')=1$, we get 
$$\forall k ,\ \|b_k(W\diag(\omega_j) - \diag(\omega_j)W)\|\leq 
2\|b_k\|_2\|W-I\|_2\|\omega_j\|_E< \sqrt{n}\Nm(\cL')^{1/n}\leq 
\lambda_1(\cL').$$
Since $M_{\cL'} = M_\cL W$, each $b_k(W\diag(\omega_j) - \diag(\omega_j)W)$ is a vector of $\cL'$, therefore 
they have to be $0$.
\end{proof}

\begin{proof}[Proof of Theorem~\ref{thm:pseudoinjectivity}]
We need to show that there are $r,\varepsilon>0$ such that 
$$
d_{G/L}(x,y)\geq r \Rightarrow |\langle f(x)|f(y)\rangle|<\varepsilon,
$$
where $d_{G/L}$ is the regular Euclidean distance in $G/L$, i.e. $d_{G/L}(x,y) = \min_{u\in L}\| x-y-u\|$. Let $\cL=f_c(x)$ be the lattice corresponding to $x$ and $\cL'=f_c(y)$ be the one corresponding to $y$. With the notations of Definition~\ref{def:aisub}, whenever $\Lambda\subsetneq\mathcal{L}$ (and $\Lambda'\subsetneq\mathcal{L}'$), we necessarily have $|\langle f(x)|f(y)\rangle|\leq 3/4$. Hence, by contraposition, we assume that $|\langle f(x)|f(y)\rangle| > 3/4$ (which implies $\Lambda = \mathcal{L}$ and $\Lambda' = \mathcal{L}'$), and we prove that this implies that $d_{G/L}(x,y)$ must be less than a certain bound $r$. 

First, Lemma~\ref{lem:1} implies that there is $\diag(a_i)_{i\leq n}$ (in the sense of the diagonal matrices discussed in Lemma~\ref{lem:diag-matrices}) such that $M_{\mathcal{L}'} = M_{\mathcal{L}}W$ for $W = e^{\diag(a_i)}$ where $\|a\|\leq \frac{1}{4\sqrt{n}|\Delta|}$. This means that the matrix distance $\dist(\cL,\cL')$ is necessarily less than $\frac{1}{4\sqrt{n}|\Delta|}$.

Next, we want to prove that if  $\dist(\cL,\cL') = \|A\|$ for some $A\in\operatorname{GL}_n(\R)$ with $M_{\mathcal{L'}} = M_{\mathcal{L}}e^A$, then $A$ is necessarily of the form $\diag(a'_i)$. We know that $\|A\|\leq \|\diag(a_i)\|\leq \frac{1}{4\sqrt{n}|\Delta|}$. Moreover, for all $A$ close to the zero matrix, the expansion of the matrix exponential tells us that 
$$
\|e^{A} - I\| = \|\sum_{k\geq 1}\frac{A^k}{k!}\| \leq \|A\|\sum_{k\geq 0}\frac{\|A\|^k}{k!}\leq (e-1)\|A\| < 2\|A\|.
$$
Hence $\|e^A-I\|< \frac{1}{2\sqrt{n}|\Delta|}$ and we can apply  Claim~\ref{claim:diag_matrix} to argue that $e^A$ is diagonal. Therefore, Since all $e^A$ with $M_{\mathcal{L'}} = M_{\mathcal{L}}e^A$ and $\dist(\cL,\cL') = \|A\|$ must be diagonal, we have that the matrix distance satisfies $\dist(\mathcal{L},\mathcal{L}') = \|a\|$ for some $a$ with $M_{\cL'} = M_{\cL}e^{\diag(a)}$ (where $\diag$ of matrices in $\R^{n\times n}$ is still understood as in Lemma~\ref{lem:diag-matrices}). 

In terms of $E$-lattices, this means that $\cL' = \phi(a^G)\cdot \cL$ where $a^G$ is an element of $G$ corresponding to $a$. To construct such an element, we first notice that $\det(e^{\diag(a)})=1$, which means that the element $x^a\in\R^{n_1}\times\C^{n_2}$, corresponding to  $e^{\diag(a)}$ satisfies $|x^a_1| = \frac{1}{\prod_{i=2}^{n_1} |x^a|_i\prod_{i=n_1+1}^{n_2}|x^a_i|^2}$. We can therefore follow the construction of elements of $G$ from $S$-units by treating $x^a\in \R^{n_1}\times\C^{n_2}$ as if it were in $U_S$ with all valuations according to primes in $S$ being $0$. (i.e. all coordinates of $a^G$ according to $\Z^s$ are set to 0). Since $M_{\phi(a^G)} = e^{\diag(a)}$ is close to the identity matrix, we notice that this construction also directly implies that the real entries of $\phi(a^G)=x^a$ are close to $1$, i.e. they are positive, and therefore all entries of $a^G$ according to $\Z_2^{n_1}$ are zero. Moreover, each diagonal block $\Xi_i$ corresponding to a complex coordinate of $\phi(a^G) = x^a$ is close to the identity block: 
$$
\Xi_i = 
\begin{pmatrix}
    \mathfrak{Re} (x^a_i) & -\mathfrak{Im}(x^a_i)\\
    \mathfrak{Im}(x^a_i) & \mathfrak{Re} (x^a_i)
\end{pmatrix}
\sim \begin{pmatrix}
    1 & 0 \\ 
    0 & 1
\end{pmatrix}.
$$
More specifically, since $\|e^{\diag(a)}-I\|<\frac{1}{2\sqrt{n}|\Delta|}$, we know that 
\begin{itemize}
    \item $|\mathfrak{Im}(x^a_i)|\leq \frac{1}{2\sqrt{n}|\Delta|}$
    \item $| \mathfrak{Re} (x^a_i) - 1|\leq \frac{1}{2\sqrt{n}|\Delta|}$.
\end{itemize}
Hence, if $\theta_i\in\R/\Z$ satisfies $x^a = |x^a|e^{2i\pi\theta_i}$, we have 
$$
|\theta_i|\leq \frac{\pi}{2}|\sin(\theta_i)| = \frac{\pi}{2}\frac{|\mathfrak{Im}(x^a_i)|}{|\mathfrak{Re} (x^a_i)|}\leq \pi |\mathfrak{Im}(x^a_i)| \leq \frac{\pi}{2\sqrt{n}|\Delta|}
$$
Let $r = \frac{\pi}{2|\Delta|}$. We have $\|a^G\|\leq r$. Hence $d_{G/L}(x,y) \leq r$. This proves by contraposition that if $d_{G/L}(x,y)\geq r$, then $|\langle f(x)|f(y)\rangle|<\varepsilon = 3/4$.

\end{proof}

\section{Lipschitz property of
  $(\textbf{y},\textbf{v})\mapsto
  |\phi(\textbf{y}) \underline{\OO}\prod_{\pg\in
    S}\pg^{-v_i}\rangle$}
\label{ssec:lipschitz}

\begin{proposition}[Lipschitz property of $f$]
There is $a > 0$ such that 
$$\left|  |f(x)\rangle - | f(y) \rangle\right| <  a\cdot d_{G/L}(x,y)$$
\end{proposition}
\begin{proof}
Let $z\in G$ such that $z = x-y-u$ where $u\in L$ is such that 
$d_{G/L}(x,y) = \|x-y-u\|$. If one of the components of $z$ according to 
$\Z_2^{n_1}$, or $\Z^s$ is non-zero, then $d_{G/L}(x,y)\geq 1$, and then 
by the triangle inequality 
$\left| \langle f(x) | f(y) \rangle\right|\leq 2\leq 2d_{G/L}(x,y)$. 
Now we assume that all components of $z$ according to 
$\Z_2^{n_1}$ and $\Z^s$ are zero. In particular, this means that 
$\mathcal{L} = (e^z)\mathcal{L'}$ where $\mathcal{L} = f_c(x)$, $\mathcal{L'}=f_c(y)$, 
and $z\in E$ correspond to the canonical mapping of the components of $z$ 
according to $\R^{n_1+n_2-1}\times (\Z_2)^{n_1}\times \left(\R/\left(\frac{1}{n^2}\Z\right)\right)^{n_2}$. Therefore, 
we have $M_{\mathcal{L}} = M_\mathcal{L'}\cdot e^{\diag(z)}$, and thus: 
\begin{align*}
d_{G/L}(x,y) &= \|z\| \geq \|\phi(z)\|= \|\diag(\phi(z))\|_2\\
             &\geq \inf\{\|A\|_2: M_{\mathcal{L}} = M_\mathcal{L'}\cdot e^{A}\}\\
             &\geq a_0 \left|  |f(x)\rangle - | f(y) \rangle\right| \ \ \text{by~\cite[Th. D.4]{EHKS14long}}
\end{align*}
Then we obtain the desired result with $a = \max\{2,1/a_0\}$. 
\end{proof}
We have demonstrated the HSP property of $f$. We will now use this to derive the HSP property of 
$\tilde{f}$ which is a function from $\tilde{G}$ to $\mathcal{H}$ obtained from $f$.  

\section{An HSP oracle on $\R^m$}\label{sec:R-oracle}

In the previous sections we described an oracle $f:G\rightarrow \mathcal{H}$ which satisfies the HSP properties of Definition~\ref{def_HSPRm} (in particular: pseudoinjectivity and Lipschitz property). We now show how to construct an oracle over $\R^m$ that hides the $S$-unit groups and that inherits the HSP properties of $f$. 
The control group $G$ can be seen as the projection of $\tilde{G} = \R^k\times \Z^l$ where 
\begin{itemize}
 \item $k = n_1 + 2n_2 -1$.
 \item $l = n_1 + s$.
\end{itemize}
We denote by $\gamma:\tilde{G}\rightarrow G$ the projection map, and by $\tilde{L}\subseteq\tilde{G}$ the 
pre-image of $L$ by $\gamma$. 

\begin{Definition}[Oracle on $\tilde{G}$]
We define $\tilde{f}:\tilde{G}\rightarrow \mathcal{H}$ by
$$
\tilde{f}(\tilde{x}) =  f\circ\gamma(\tilde{x}).
$$
\end{Definition}
We have the following diagram: 

\[\begin{tikzcd}
\R^m & \tilde{G} = \R^k\times \Z^l \arrow[l]
                                   \arrow[r,"\gamma"]
                                   \arrow[dl,"\tilde{f}"]
                                   \arrow[d,"\tilde{\alpha}"]
    & G = \R^{n_1+n_2-1}\times \Z_2^{n_1}\times \left(\R/\Z\right)^{n_2}\times \Z^s \arrow[d,"\alpha"]
                                                                         \arrow[dr,"f = f_q\circ f_c"]
    & \\
\mathcal{H} & \tilde{G}/\tilde{L} \arrow[r,"\pi"] & G/L & \mathcal{H}
\end{tikzcd}\]

We proceed by first showing that $\tilde{f}$ satisfies the HSP properties, and then we use techniques from~\cite[Th. 6.1]{STOC2014} and~\cite[Sec. F]{EHKS14long} to derive an oracle on $\R^m$ that satisfies the HSP properties and that hides the $S$-unit group.  

\subsection{HSP properties of the oracle on $\tilde{G}$}

In this section, we show that $\tilde{f}$ is an a,$r,\varepsilon$-oracle on $\tilde{G}$. Following the framework of~\cite{STOC2014,EHKS14long}, we use the following distance on $\tilde{G}/\tilde{L}$.

\begin{Definition}[Distance on $\tilde{G}/\tilde{L}$]
 Let $\tilde{x},\tilde{y}\in\tilde{G}$. We define $d_{\tilde{G}}(x,y) = \|x-y\|$ if $x-y$ does not have any non-zero components on $\Z^l$, and  $d_{\tilde{G}}(x,y) = \infty$ otherwise. Then
 $$
 d_{\tilde{G}/\tilde{L}} = \inf_{\tilde{u}\in\tilde{L}} d_{\tilde{G}}(x , y+u).
 $$
\end{Definition}

\begin{proposition}[Lipschitz property of $\tilde{f}$]
Assume $f$ is an $a,r,\varepsilon$-oracle. Then
$$
\forall \tilde{x},\tilde{y}\in\tilde{G},\ \| | \tilde{f}(\tilde{x})\rangle 
- | \tilde{f}(\tilde{y})\rangle\| \leq 
a\cdot d_{\tilde{G}/\tilde{L}}(\tilde{x},\tilde{y}).
$$
\end{proposition}
\begin{proof}
Suppose $ d_{\tilde{G}/\tilde{L}}=\infty$, then the inequality holds trivially. Otherise, 
Let $\tilde{u}\in \tilde{L}$ such that 
$d_{\tilde{G}/\tilde{L}}(\tilde{x},\tilde{y}) = \|\tilde{x}-\tilde{y}-\tilde{u}\|$. In particular, all coordinates of $\tilde{z}:= \tilde{x}-\tilde{y}-\tilde{u}$ 
with respect to $\Z^l$ are $0$. Let $u=\gamma(\tilde{u})$, $x=\gamma(\tilde{x})$, and $y=\gamma(\tilde{y})$. 
We have 
$$
\|\tilde{x}-\tilde{y}-\tilde{u}\| 
\geq \|x - y - u\| \geq \min_{u\in L} \|x-y-u\| = d_{G/L}(x,y).
$$
Hence 
$a\cdot d_{\tilde{G}/\tilde{L}}(\tilde{x},\tilde{y}) \geq a\cdot d_{G/L}(x,y) 
\geq \| | \tilde{f}(\tilde{x})\rangle 
- | \tilde{f}(\tilde{y})\rangle\|$
\end{proof}

\begin{proposition}[Pseudoinjectivity of $\tilde{f}$]
Assume $f$ is an $a,r,\varepsilon$-oracle for $r\ll 1$. Then
$$d_{\tilde{G}/\tilde{L}}(\tilde{x},\tilde{y})\geq r \Rightarrow \left| \langle \tilde{f}(\tilde{x}) | \tilde{f}(\tilde{y}) \rangle\right| < \varepsilon$$
\end{proposition}

\begin{proof}
Let $x = \gamma(\tilde{x})$, $y = \gamma(\tilde{y})$, 
and $u\in L$ such that $d_{G/L}(x,y) = \|x-y-u\|$. 
If $z = x-y-u$ has no component on $\Z_2^{n_1}$ or $\Z^s$, then $d_{\tilde{G}/\tilde{L}}(\tilde{x},\tilde{y}) = d_{G/L}(x,y)$ and 
therefore, if $d_{\tilde{G}/\tilde{L}}(\tilde{x},\tilde{y}) \geq r$, then 
$$
|\langle \tilde{f}(\tilde{x}) | \tilde{f}(\tilde{y})\rangle|
= |\langle f(x) | f(y)  \rangle| \leq \varepsilon.
$$

On the other hand, if for such a $u$, we have components on $\Z_2^{n_1}$ or 
$\Z^s$, then either $d_{\tilde{G}/\tilde{L}}(\tilde{x},\tilde{y}) = d_{\tilde{G}}(\tilde{x},\tilde{y}+\tilde{u})=\infty$, 
where $\tilde{u}\in\tilde{L}$ is the corresponding preimage, or $d_{\tilde{G}/\tilde{L}}(\tilde{x},\tilde{y})\geq 1$. So 
we only know that $d_{\tilde{G}/\tilde{L}}(\tilde{x},\tilde{y}) \geq d_{G/L}(x,y)$ in this case. However, 
we also have that $d_{G/L}(x,y)\geq 1$ because of the integer components. Since $r\ll 1$ we necessarily have $d_{G/L}(x,y)\geq r$, and therefore $
|\langle \tilde{f}(\tilde{x}) | \tilde{f}(\tilde{y})\rangle|
= |\langle f(x) | f(y)  \rangle| \leq \varepsilon.
$
\end{proof}

\subsection{An HSP oracle on $\R^m$}

Assume we have an $\tilde{a},\tilde{r},\tilde{\varepsilon}$-oracle $\tilde{f}$ that hides $U_S$ on $\tilde{G} = \R^{k}\times \Z^l$. Following~\cite[Th. 6.1]{STOC2014} and~\cite[Sec. F]{EHKS14long} we derive an oracle $g:\R^m\rightarrow\mathcal{H}$ for $m = k+l$ defined by 
$$
\ket{g(\vect{x},y_1,\ldots,y_l)} = 
\sum_{z_1,\ldots,z_l\in\{0,1\}}\left( \bigotimes_{j=1}^l\ket{\psi(y_j,z_j} \right)
\otimes\ket{\tilde{f}(\vect{x},s(y_1,z_1),\ldots,s(y_l,z_l)},
$$
where $s(y,z) = \lfloor y/\lambda\rfloor+z$, and 
$\ket{\psi(y,z)} = \cos(\frac{\pi t}{2})\ket{\str{t}}$ with $t = y/\lambda - s(y,z)$.

\begin{theorem}[Theorem~6.1 of~\cite{STOC2014}]
   If $\tilde{f}$ is an   $\tilde{a},\tilde{r},\tilde{\varepsilon}$-oracle, then $g$ is an $a',r',\varepsilon'$-oracle with the following identities: 
   \begin{align*}
       a'^2 &= \tilde{a}^2 + l\left(\frac{\pi}{2\nu\lambda}(1+\nu)\right)^2\\
       r'^2 &= \tilde{r}^2 + l(2\nu\lambda)^2\\
       \varepsilon' &= \tilde{\varepsilon}.
   \end{align*}
\end{theorem}

\subsection{Concrete parameters for the $\R$-grid}

Finally, we need to bound the first minima and the fundamental volume of the lattice of $S$-units. In 
the following, we show that these values have polynomial size with respect to the input. 
To bound the first minima of $U_S\subseteq G$ and the volume of $G/U_S$ (which are preserved by the embedding of 
$U_S$ into $\tilde{G}$), we rely on an analogue of 
Dirichlet unit theorem that applies to $S$-units. The classical results are known for the case where 
the lattice of $S$-units is embedded in $\R^{r+|S|}$ (where $r$ is the rank of the unit group of $\OO$) 
via the logarithm embedding 
$$\Log(\alpha) := \left( \log(|\alpha|_1),\cdots,\log(|\alpha|_r),\log(|\alpha|_{\pg_1}),\cdots, 
\log(|\alpha|_{\pg_{|S|}}\right),$$
where $|\alpha|_j := |\sigma_j(\alpha)|$ and $|\alpha|_{\pg_j}:= p_j^{-e_jv_{\pg_j}(\alpha)}$. In this case, 
we know from~\cite[Lem. 2]{Hadju_S-unit} that $\|\Log(\alpha)\|_{\infty} \geq \frac{\log(n)}{6n^4}$ where $\|v\|_{\infty}$ denote the 
usual infinity norm on the vector v, and 
$$\Vol\left(\R^{r+|S|}/\Log(U_S)\right) \leq \left( 300 \log(P) \sqrt{|\Delta|} 
\left( \frac{e}{2} \log(|\Delta|) \right)^{n-1}\right)^{|S|+r-\frac{n}{2}},$$
where $P = \max_j\Nm(\pg_j)$ (see~\cite[Sec. 2]{Hadju_S-unit}). 


\begin{proposition}
The first minima of $U_S\subseteq G$ satisfies $\lambda_1(U_S)\geq \frac{\log(n)}{6n^4}$ where the norm on 
elements of $G$ is defined by 
$$\|(z,v_1,\cdots,v_{|S|})\| := \sqrt{\sum_j z_j^2} + \sum_j |v_j|e_j\log(p_j).$$
Moreover, the volume of the lattice of $S$-units satisfies 
$$\Vol(G/U_S) \leq \frac{2^{n_1}}{\log(2)^{|S|}} \left( 300 \log(P) \sqrt{|\Delta|} 
\left( \frac{e}{2} \log(|\Delta|) \right)^{n-1}\right)^{|S|+r-\frac{n}{2}},$$
where $P = \max_j\Nm(\pg_j)$.
\end{proposition}

\begin{proof}
Let $((z_j),(v_k))\in G$ corresponding to an $S$-unit $\alpha$. We immediately see that 
$$\|((z_j),(v_k))\| \geq \|\Log(\alpha)\|_{\infty},$$ 
which proves the lower bound on 
$\lambda_1(U_S)$. 

To compute an upper bound on the volume of $G/U_S$, we follow the same approach as~\cite{EHKS14long}. First, we consider the exact sequence 
$0\rightarrow \Z^{n_1}\times(\R/\Z)^{n_2}
\rightarrow G\rightarrow\R^{m_1+n_2-1}\times\Z^s\rightarrow 0$. Let $\mu(K)$ be the group of torsion units, and $L_S\subseteq \R^{n_1+n_2-1}\times\Z^s$ be the rank-$n_1+n_2+s-1$-lattice that is the projection of $U_S$. Then we have the exact sequence
$$
0\rightarrow (\Z^{n_1}\times(\R/\Z)^{n_2})/\mu(K)
\rightarrow G/U_S\rightarrow
(\R^{n_1+n_2-1}\times\Z^s)/L_S.
$$
Hence $\Vol(G/U_S) = \Vol\left((\Z^{n_1}\times(\R/\Z)^{n_2})/\mu(K)\right) \Vol\left( (\R^{n_1+n_2-1}\times\Z^s)/L_S\right)$

The volume of $ (\R^{n_1+n_2-1}\times\Z^s)/L_S$ is equal to the absolute value of determinant of the 
matrix of a basis of $L_S$. Let $(\alpha_j)_{j\leq r+|S|}$ be a minimal generating set for $U_S/\mu(K)$. Its matrix $M$ with 
respect to the embedding in $\R^{r}\times Z^{|S|}$ is related to the matrix $M':=(\Log(\alpha_j))$ by the 
relation $M = D\cdot M'$ where

\[ D= \left( \begin{array}{cccccc}
1 &   (0)    &     &       &         &  \\
    & \ddots &     &       &         &  \\
   &    (0)     & 1 &       &         &   \\
   &         &     &   1/e_1\log(p_1)   &         & (0)  \\
   &         &     &       &  \ddots &   \\
   &         &     &   (0)    &         & 1/e_{|S|}\log(p_{|S|})
\end{array} \right).\]
We therefore have 
\begin{align*}
\Vol\left( (\R^{n_1+n_2-1}\times\Z^s)/L_S\right) &= \det(M) = \prod_j \frac{1}{e_j\log(p_j)} \det(M')\\
&= \prod_j \frac{1}{e_j\log(p_j)}\Vol(R^{r+|S|}/\Log(U_S)).
\end{align*}
Additionally, we have 
$\Vol\left((\Z^{n_1}\times(\R/\Z)^{n_2})/\mu(K)\right) = \frac{2^{n_1}}{|\mu(K)|}$, therefore, by using the upper bound on  $\Vol(R^{r+|S|}/\Log(U_S))$, we get
$$\Vol(G/U_S) \leq \frac{2^{n_1}}{\log(2)^{|S|}} \left( 300 \log(P) \sqrt{|\Delta|} 
\left( \frac{e}{2} \log(|\Delta|) \right)^{n-1}\right)^{|S|+r-\frac{n}{2}}.$$
\end{proof}

We are left to invoke the efficient HSP algorithm on $\R^m$. We recall its performance below. 
\begin{theorem}[Theorem 6.2~\cite{EHKS14long}]
Let $f$ be an HSP instance with parameters $((a,r,\varepsilon), \lambda, d)$ such that $r \le \frac{1}{36}{\lambda}$ and $\varepsilon\le 1/2$. Let $0<\eta\le\lambda_1(L^*)$ and $0<\mu<1$ be a precision parameter and error parameter respectively. There is a quantum algorithm that makes $K=O\left(\log d + \log(1/\mu) + m\log a \right)$ oracle calls to $f$ and generates $K$ vectors $\tilde u^{(1)},\dots \tilde u ^{(s)}\in\R^{m}$ having the following properties except with error probability at most $\mu$:
\begin{enumerate}
\item\; $\|\tilde u^{(i)}\|\le\frac{3a}{\pi}$ for $i=1,\dots,K$;
\item\; For each $i$, $\|\tilde u^{(i)}-u^{(i)}\|\le\eta$ for some $u^{(i)}\in L^*$;
\item\; $u^{(1)},\dots,u^{(K)}$ generate $L^*$.
\end{enumerate}
The quantum circuit uses $O\bigl(Km\bigl(\log\frac{aK}{\eta\mu}\bigr)^2\bigr)$ quantum gates on $O\bigl(m\log\frac{aK}{\eta\mu}\bigr)$ qubits.
\label{hspr_complexity}
\end{theorem}

The above allows us to derive precise polynomial dependencies in the input parameters. For this, one needs to specify the cost of running the LLL algorithm in superposition. Generic transformations exist to turn a classical algorthm into a reversible quantum computation~\cite{Ben89}. To facilitate a comparison with the recent prepring of de Boer and Felderhoff, we specialize our result to the case of the quantum LLL algorithm of~\cite[Eq~7,8]{TiepeltS19}.

\begin{corollary}
Let $K$ be an number field of degree $n$, and discriminant $\Delta$. Let $S$ be a set of prime ideals of $K$. The gate cost of each call to the oracle to compute the $S$-unit group is in   
$$
\tilde{O}\left( (n+|S|)^{5.5}(n+\log|\Delta|)^{5.5} n^{17.5}(n^2+\log|\Delta|)^{3.5}\right).
$$
Meanwhile, the qubit requirement of the oracle is in 
$$
\tilde{O}\left( (n+|S|)^{2.5}(n+\log|\Delta|)^{2.5} n^{9.5}(n^2+\log|\Delta|)^{3.5}\right).
$$
\end{corollary}

\begin{proof}
To use Theorem~\ref{th:classical-oracle_cost}, we need a bound on the bit size of the entries of the vectors given as input to the oracle. It is given by the qubit requirements of the quantum oracle, i.e. it is in 
$$
\tilde{O}\left( m ( \log(a) + \log(K) + \log(\frac{1}{\eta}) + \log(\frac{1}{\mu})\right).$$
With a precision of $2^{-O(d)}|\Delta|^{-O(1)}$, we obtain (using the notation of Theorem~\ref{th:classical-oracle_cost})
$$
\log(v),\log(r)\in\tilde{O}\left( (n+|S|)(n+\log|\Delta|)\right).
$$
This means that the bound $\beta$ on the bit size of the entries of the vectors given as input to the LLL algorithm satisfies 
$$\beta\in\tilde{O}\left( n(n^2+\log|\Delta|)(n+\log|\Delta|)(n+|S|)\right).$$
Using again $C_{\text{gates}}(k,\beta)$ and  $C_{\text{qubits}}(k,\beta)$ to denote the gate cost (resp. the qubit cost) of running LLL in superposition, we have therefore a ``classical oracle'' gate cost in 
$$
\tilde{O}\left( (n+|S|)^2(n+\log|\Delta|)^2 C_{\text{gate}}(n^2,\beta)\right)
$$
Likewise, the qubit cost of the classical oracle is in 
$$
\tilde{O}\left( (n+|S|)(n+\log|\Delta|) C_{\text{qubit}}(n^2,\beta)\right)
$$
If, similarly to~\cite{BoerF25}, we use the methods of~\cite[Eq~7,8]{TiepeltS19}, then we can assume that 
$C_{\text{gates}}(k,\beta) = O(k^7\beta^{3.5})$ and  $C_{\text{qubits}}(k,\beta) = O(k^4\beta^{1.5})$. The gate cost of our classical oracle is then in 
$$
\tilde{O}\left( (n+|S|)^2(n+\log|\Delta|)^2 n^{14} \left(n(n^2+\log|\Delta|)(n+\log|\Delta|)(n+|S|)\right)^{3.5}\right).
$$
Meanwhile, the qubit cost of the oracle is then in 
$$
\tilde{O}\left( (n+|S|)(n+\log|\Delta|) n^{8} \left(n(n^2+\log|\Delta|)(n+\log|\Delta|)(n+|S|)\right)^{1.5}\right).
$$
The cost of the calculation of the quantum encoding of the output of the ``classical oracle'' is that of a matrix multiplication, which is negligible compared to the computation of the basis of the ideal itself. 
\end{proof}

\section{Applications to other number theory problems}\label{sec:applications}

\subsection{Recover an exact representation of the $S$-units}

The solution of \hsp{} is given to 
us as approximations of generators of the hidden subgroup. For many applications, an exact (and polynomially 
bounded) representation is preferable. Therefore, we process the solutions to the \sus{} problem classically to produce a compact 
representation of the generators of the $S$-unit group. 
\begin{Definition}[Compact representation]
Let $l>0$ be a constant, a compact 
representation of $\alpha\in\OO$ with respect to the integral basis $(\omega_j)_{j\leq n}$ of $\OO$ 
is a set of exact representations of polynomial size algebraic numbers $\gamma_j$ satisfying 
$\alpha = \gamma_0\gamma_1^{l}\cdots\gamma_k^{l^k},$ 
where $k$ is polynomial in the size of the input.
\end{Definition}
Biasse and Fieker~\cite[Sec. 5]{ANTS_XI} described an efficient 
method based on~\cite[Alg. 7.53]{fieker_notes} to classically compute a compact representation of 
an algebraic number 
in polynomial time. These methods rely on the knowledge of an exact representation of the algebraic 
number we wish to represent (which is not the case here). A modification of~\cite[Alg. 7.53]{fieker_notes} 
using the approximation of the vector corresponding to an algebraic number yields 
a compact representation of that number. 

Our algorithm for the compact representation of an $S$-unit takes as input $l>0$ and a 
rational approximation (to an arbitrary polynomial 
precision $q$) of a vector of the form 
$$(\log(|\alpha|_1,\cdots ,\log(|\alpha|_{n_1+n_2}),\theta,v_{\pg_1}(\alpha),\cdots,v_{\pg_{|S|}}(\alpha)),$$
where $\alpha$ is an $S$-unit. We can assume that $\prod_{j}\pg_j^{v_{\pg_j}(\alpha)} \subseteq\OO$. If not, 
we replace each $\log(|\alpha|_j)$ by 
$$\left( \sum_{v_{\pg_k}(\alpha) < 0} |v_{\pg_k}(\alpha)|e_k\log(p_k)\right) + \log(|\alpha|_j)$$ 
(where $\Nm(\pg_k) = p_k^{e_k}$), thus 
calculating a compact representation of $\alpha\prod_{v_{\pg_k}(\alpha) < 0} p_k^{|v_{\pg_k}(\alpha)|e_k}$. 
From that, we can easily derive a compact representation of $\alpha$. Then, we choose $k_0$ minimal such that 
$\frac{\log(|\alpha|_j}{l^{k_0}} \leq \log(|\Delta|)$, initiate an ideal $I$ to 
$\prod_{j}\pg_j^{\left\lfloor v_{\pg_j}(\alpha)/l^k\right\rfloor}$, and we compute rational approximations $v_j$ of 
$\left(|\alpha|_j\right)^{1/l^k}$. Then at each step, $I$ is replaced by $I^l$ and we compute an 
LLL-reduced element $\delta_k$ of the ideal $C\subseteq\OO$ such that $I^{-1} = \frac{1}{d_k} C$ for 
the scaled $T_2$ norm $T_{2, (v_j)_j}(\delta) := \sum |\delta|^2_i \frac{v_j^2}{v^2}$ where 
$v := \sqrt[n]{\prod v_j}$. The ideal $I$ is then replaced by $\beta_k I$ where $\beta_k := \frac{\delta_k}{d_k}$, 
and $v_j\leftarrow v_j\cdot|\beta_k|_j$. At every step $k$ from $k_0$ to $0$, we know that 
\begin{itemize}
 \item $\beta_k$ has polynomial size,
 \item $\beta_{k_0}^{l^{k}}\cdots\beta_k\left(\alpha\right)^{\frac{1}{l^{k_0-k}}}$ has polynomial size,
 \item $I\subseteq \OO$ and has polynomial size (i.e. $\log(I)$ is polynomial),
 \item $\prod_j v_j \geq \Nm(I)\geq 1$. 
\end{itemize}
At the end of this process, we have polynomial size algebraic numbers $(\beta_j)_{j\leq k_0}$ such that 
$\beta_{-1} := \alpha\prod_k \beta_k^{l^k}$ has polynomial size. Finding $\beta_{-1}$ is the main difference 
between our approach and that of~\cite[Sec. 5]{ANTS_XI} and~\cite[Sec. 7]{fieker_notes} since we have no 
exact representation for $\alpha$. 
We find the minimal $d>0$ such that 
$\beta := d\beta_{-1}\in \OO$ and from approximates of the $\log(|\beta_k|_j)$, $\log(|\alpha|)_j$, 
and the phase vector of each of the corresponding algebraic numbers, we 
find a rational approximation $\widehat{\beta}\in\R^m$ under the rule~\eqref{eq:map_Rm} with a 
polynomial number of bits of precision. Likewise, we can get approximations $\widehat{\omega_j}\in\R^m$ 
of the integral basis vectors $\omega_j$, and solve the linear system (over the rationals) 
$\widehat{\beta} := \sum_j \frac{b_j}{c_j}\widehat{\omega_j}$. The nearest lattice point 
$\sum_j a_j\widehat{\omega_j}$ in $\sum_j\Z\widehat{\omega_j}$ 
can be retrieved if the precision is larger than $n$ by using Babai's algorithm~\cite{barbai}. Then 
we know that $\beta = \sum_j a_j\omega_j$, and 
$$\alpha = \frac{\beta_{-1}}{\beta_0}\left(\frac{1}{\beta_1}\right)^ l\cdots
\left(\frac{1}{\beta_{k_0}}\right)^{l^{k_0}}.$$

\begin{algorithm}[ht]
\caption{Compact representation}
\begin{algorithmic}[1]\label{alg:compact}
\REQUIRE Rational approximations of $\log(|\alpha|_j)$, phase vector of $\alpha$ and $v_j\geq 0$ such that 
$(\alpha) = \prod_j\pg_j^{v_j}$, $l>0$, and approximations $\widehat{\omega_j}$ of an LLL-reduced integral basis of $\OO$. 
\ENSURE Exact representation of $\gamma_0,\cdots,\gamma_{k_0}$ such that 
$\alpha = \prod_k \gamma_k^{l^k}$.
\STATE $I\leftarrow\prod_{j}\pg_j^{\left\lfloor v_{\pg_j}(\alpha)/l^k\right\rfloor}$.
\STATE Let $k_j$ minimal such that $\frac 1 {l^k} \log|\beta_j|_i\le \log\Delta$, 
$v_j\leftarrow\exp(l^{-k}\log|\alpha|_j)$
\FOR {$0\leq k\le k_0$}
\STATE $B \leftarrow I^l$, $(w_j)_j \leftarrow (v_j^l)_j$.
\STATE $w \leftarrow \sqrt[n]{\prod w_j}$ and $d_k\in\Z_{>0}$ such that $B^{-1}=\frac{1}{d_k}C$ for $C\subseteq\OO$. 
\STATE Let $\delta$ be a 1st LLL-basis element of $C$ with respect to $T_{2, (w_j/w)_j}(\delta) := \sum |\delta|^2_i \frac{w_j^2}{w^2}$.
\STATE $\beta_k\leftarrow \frac{\delta}{d_k}$, $I \leftarrow B\beta_k$, $(v_j)_{j\leq r+1} \leftarrow  (w_j\cdot|\beta_k|_j)_{j\leq r+1}$.
\ENDFOR
\STATE Let $\beta_{-1} = \alpha\cdot \prod_k \beta_k^{l^k}$
\STATE Find an approximation $\widehat{\beta}\in\R^m$ of $d\beta_{-1}$ where $d\in\Z_{>0}$ is minimal such that $d\beta_{-1}\in\OO$.
\STATE Find $(a_j)_{j\leq n}$ such that $\sum_j a_j\widehat{\omega_j}$ is the closest vector to $\widehat{\beta}$ in $\sum_j\Z\widehat{\omega_j}$. 
\STATE $\beta_{-1} \leftarrow \frac{1}{d}\sum_j a_j\omega_j$. 
\RETURN $\frac{\beta_{-1}}{\beta_0},\frac{1}{\beta_1},\cdots,\frac{1}{\beta_{k_0}}$. 
\end{algorithmic}
\end{algorithm}

\begin{proposition}
Algorithm~\ref{alg:compact} is correct and returns a compact representation of the input $\alpha$ in polynomial time. 
\end{proposition}

\begin{proof}
The invariant properties on the size of the elements are deduced in the same way as in the proof of~\cite[Prop. 5.1]{ANTS_XI}. 
The only important different is the way we compute an exact representation of $\beta_{-1}$. Barbai's algorithm allows us 
to find in polynomial time a lattice element $\tilde{\beta}$ in $\widehat{\mathcal{L}}:=\sum_j\Z\widehat{\omega_j}$ 
such that $d(\widehat{\beta},\tilde{\beta})\leq 2^n d(\widehat{\beta},\widehat{\mathcal{L}})$. If the precision is 
larger than $n$, then the coefficients of $\tilde{\beta}$ on the basis $\widehat{\omega}_j$ are those of 
$\beta = d\beta_{-1}$ on the integral basis $\omega_j$ of $\OO$.  
\end{proof}

\subsection{Computation of class groups}

Let $\mathcal{B} = \{ \p_1,\cdots,\p_N\}$ be a set of invertible prime ideals of an order 
$\OO$ whose classes generate $\Cl(\OO)$. 
We have a surjective morphism 
$$\begin{CD}
\Z^N @>{\varphi}>> \mathcal{I} @>{\pi}>> \Cl(\OO)\\
(e_1, \ldots, e_N) @>>> \prod_i\p_i^{e_i} @>>> \prod_i[\p_i]^{e_i}
\end{CD},$$
and the class group  $\Cl(\OO)$ is isomorphic to $\Z^N/\ker(\pi\circ\varphi).$ Therefore, computing the class group boils down to 
computing $\ker(\pi\circ\varphi)$, which is the lattice of $(e_1,...,e_N)\in\Z^N$ such that 
$\p_1^{e_1},\hdots , \p_N^{e_N} = (\alpha)$ for some $\alpha\in K$. These $\alpha$ are $S$-units for $S = \mathcal{B}$, and the exponent vectors 
of a generating set of $U_S$ give us a generating set for $\ker(\pi\circ\varphi)$ from which we derive $\Cl(\OO)$ 

 The best unconditional bounds on $|S|$ are exponential in $\log(|\Delta|)$. As the complexity 
of the computation of the $|S|$-unit group is polynomial in $|S|$, we cannot achieve a polynomial complexity 
unconditionally that way. However, under the Generalized Riemann Hypothesis (GRH), the classes of all prime 
ideals of $\OO$ of norm up to $48\log(|\Delta|)^2$ generate $\Cl(\OO)$. The size of 
$S := \{ \pg\subseteq\OO \text{ prime } \mid \Nm(\pg)\leq 48\log(|\Delta|)^2\}$ is polynomial in $\log(|\Delta|)$, 
and calculating the corresponding $S$-unit group is therefore polynomial in $n$ and $\log(|\Delta|)$. 

\begin{algorithm}[ht]
\caption{Ideal class group of $\OO$}
\begin{algorithmic}[1]\label{alg:class_group}
\REQUIRE $\OO$ 
\ENSURE $d_1,\cdots,d_n$ such that $\OO\simeq \Z/d_1\Z\oplus\cdots\oplus \Z/d_n\Z$. 
\STATE $S\leftarrow \{\pg\subseteq\text{ prime }\mid \Nm(\pg)\leq 48\log(|\Delta|)^2\}$.
\STATE Compute the $S$-unit group $U_S$.
\STATE Let $(\alpha_j,v_{j,1},\cdots,v_{j,|S|})_{j\leq r + |S|}$ be the generating set for $U_S$ computed.
\STATE $\diag(d_1,\cdots,d_n)\leftarrow $ Smith Normal Form of $M = (v_{j,k})$.
\RETURN $d_1,\cdots,d_n$. 
\end{algorithmic}
\end{algorithm}

\begin{proposition}
Under the Generalized Riemann Hypothesis, Algorithm~\ref{alg:class_group} is correct and runs in 
polynomial time.
\end{proposition}

 Our work also has direct applications in computational number theory. Indeed, the $S$-unit group is 
a central object that can be used in a lot of algorithms. It usually is computed together 
with the so-called $S$-class group, which is the quotient of the group of ideals in the ring of $S$-integers 
by the subgroup of principal ideals. The $S$-class group can easily be derived from the ideal class group and 
an oracle for the PIP by quotienting the class group by extra relations. A description of this method 
can be found in Simon's PhD thesis~\cite[Chap. 1]{Simon}.

Another direct consequence of our work is that it directly implies a polynomial time algorithm for 
computing the relative class group and the relative unit group of an arbitrary extension of number fields. Algorithms 
for these tasks are already known~\cite{Cohen2}[Ch. 7], but their run time is exponential in the 
degree of the fields. As for the $S$-class group, they also consist 
of using a complete set of relations for the ideal class group and of enriching it with new relations that 
are obtained by solving instances of the PIP.

\subsection{Resolution of the principal ideal problem}

Let $\ag\subseteq\OO$ be an ideal of $\OO$. We want our algorithm to run in polynomial time 
in the size of the input, that is $\log|\Delta|$, $n$, and $\log(\Nm(\ag))$ (which quantifies the size of $\ag$). 
The ideal $\ag$ is principal if and only if $\ag = (\alpha)$ for $\alpha$ an $S$-unit where $S$ is the set of 
prime divisors of $\ag$. We calculate a generating set for the $S$-units, which gives us a generating set for 
all the possible principal ideals only divisible by elements of $S$. The resolution of a linear system tells us 
if $\ag$ belongs to this set, and if so, what is its generator.
%
%
%
\begin{algorithm}[ht]
\caption{Principal ideal problem}
\begin{algorithmic}[1]\label{alg:PIP}
\REQUIRE $\OO$ and an ideal $\ag\subseteq\OO$. 
\ENSURE Decide if $\ag$ is principal and if so a compact representation of a generator $\alpha$.
\STATE Factor $\ag$, let $S = \{\pg_1,\cdots,\pg_k\}$ be the divisors of $\ag$.
\STATE Compute the $S$-unit group $U_S = \mu\times\langle\varepsilon_1\rangle\times \cdots\times\langle\varepsilon_{r+|S|}\rangle$.
\STATE Let $M = (m_{i,j})$ such that $\varepsilon_i = \prod_j\pg_j^{m_{i,j}}$.
\STATE Solve $XM = \textbf{a}$ where $\ag = \prod_i\pg_i^{a_i}$.
\RETURN compact representation of $\prod_i\varepsilon_i^{x_i}$ or ``not principal'' if the system has no solution.
\end{algorithmic}
\end{algorithm}

\subsection{Ideal class decomposition in $\Cl(\OO)$}

Under the GRH, the set of prime ideals
$$S := \{\pg\subseteq\OO \text{ prime } \mid \Nm(\pg)\leq 48\log(|\Delta|)^2\}\cup \{ \pg\subseteq\OO \text{ prime } \mid \pg\mid \ag\}$$
generate the ideal class group. Ideal class decomposition consists of finding exponents $x_1,\ldots,x_s$ and $\alpha\in K$ such that 
$$
\ag = (\alpha)\pg_1^{x_1}\ldots \pg_s^{x_s}.
$$
 We want our algorithm to run in polynomial time 
in the size of the input, that is $\log|\Delta|$, $n$, and $\log(\Nm(\ag))$ (which quantifies the size of $\ag$). Our strategy is the following:
\begin{enumerate}
 \item Decompose $\ag$ as a product of prime ideals $\ag = \prod \qg$. 
 \item For each $\qg_j\notin\mathcal{B},j\leq k$ in the decomposition of $\ag$, find $\beta_k\in K$ such that 
 $\qg = (\beta_k) \cdot \prod_{\pg_j\in S}\pg^{x_{j,k}}$.
 \item Deduce $\textbf{v}\in\Z^{N}$ such that  $\ag = \prod_k(\beta_k) \cdot \prod_{\pg_j\in\mathcal{B}} \pg^{v_j}$.  
 \end{enumerate}

deciding if an input ideal $\ag\subseteq\OO$ is principal, and if so, 
compute an element $\alpha\in \OO$ such that $\ag=(\alpha)$.
The first step consist of finding the prime ideal decomposition of $\ag$. Then we define $S$ by
$$S := \{\pg\subseteq\OO \text{ prime } \mid \Nm(\pg)\leq 48\log(|\Delta|)^2\}\cup \{ \pg\subseteq\OO \text{ prime } \mid \pg\mid \ag\},$$
and compute the $S$-unit group. Then we deduce the solution to the principal ideal problem by performing a linear 
algebra step on the matrix of the valuations, as described in Section~\ref{subsec:high_level_PIP}.

\begin{algorithm}[ht]
\caption{Ideal class decomposition}
\begin{algorithmic}[1]\label{alg:DLP}
\REQUIRE $\OO$ and an ideal $\ag\subseteq\OO$. 
\ENSURE Decide if $\ag$ is principal and if so a compact representation of a generator $\alpha$.
\STATE Factor $\ag$.
\STATE $S\leftarrow \{\pg\subseteq\text{ prime }\mid \Nm(\pg)\leq 48\log(|\Delta|)^2\}$.
\STATE $S\leftarrow S\cup \{ \pg\subseteq\OO \text{ prime } \mid \pg\mid \ag\}$.
\STATE $\textbf{v}\leftarrow$ vector of valuations of $\ag$ according to $S$.
\STATE Compute the $S$-unit group $U_S$.
\STATE Let $(\alpha_j,v_{j,1},\cdots,v_{j,|S|})_{j\leq r + |S|}$ be the generating set for $U_S$ computed.
\STATE Compute a compact representation of the $\alpha_j$.
\STATE Find $U\in GL_{r+|S|}(\Z)$ and $H$ such that $U\left(\frac{H|0}{B|I}\right)$ is the HNF of $(v_{j,k})$ and $I=I_m$.
\STATE $\beta_j \rightarrow \prod_k \alpha_k^{U_{j,k}}$ in compact representation for $j\leq r+|S|$.
\STATE $\textbf{v} \rightarrow\textbf{v}_1 + B\textbf{v}_2$ where $\textbf{v} = (\textbf{v}_1,\textbf{v}_2)$.  
\RETURN $\prod_k(\beta_k),\vect{v}$
\end{algorithmic}
\end{algorithm}

\begin{proposition}
Under the Generalized Riemann Hypothesis, Algorithm~\ref{alg:DLP} is correct and runs in polynomial time.
\end{proposition}

\subsection{Ray class groups}

Our algorithms also directly imply a quantum algorithm for computing the ray class group of an 
arbitrary number field. 
The computation of the ray class group is an essential task in computational class field theory, and both 
classical and quantum algorithms have been described to solve this task. A classical method due to Cohen can 
be found in~\cite{Cohen2}[3.2] and has an exponential run time with respect to the degree (but runs in subexponential 
time for classes of fixed degree number fields).  A quantum algorithm was described by Eisentr\"{a}ger and 
Hallgren~\cite{Hallgren10} with a polynomial run time in classes of fixed degree number fields. As for the 
afortmentioned tasks, computing the ray class group essentially relies on subroutines for computing the 
ideal class group and solving the PIP, for which we provide polynomial time algorithms in arbitrary number 
fields. It also relies on algorithms for factoring ideals (which can be easily derived from Shor's factoring 
algorithm), and efficient methods for solving the discrete logarithm problem (which is also a well known 
consequence of Shor's work~\cite{Sho97}). 

\subsection{Norm equations}

Finally, our work allows us to describe polynomial time algorithms for solving relative 
norm equations of the form $\Nm_{L/K}(x) = \theta$ where $L/K$ is an arbitrary Galois extension. 
Norm equations are an important example of Diophantine equations which are a major topic in 
number theory. The resolution of the Pell equation (for which there is a quantum algorithm~\cite{Hal07}) 
can be seen as a special case where $L = \Q(\sqrt{\Delta})$, $K = \Q$ and $\theta = 1$ (when we restrict our attention 
to integer solutions). Solving norm equations in general is an important task in computational number theory. A classical method 
was described by Simon~\cite{Simon} (based on the work of Fieker~\cite{Fieker_phd} for Galois extensions) 
that solves general extensions in exponential time in the degree of the fields. For the Galois case, 
it simply uses the knowledge of the $S$-unit group and the relative class group, which we can provide 
in polynomial time for number fields of arbitrary degree. However, the general method uses the Galois 
closure, whose degree can be exponential in the degree of the field, thus restricting the direct 
application of our work to arbitrary Galois extensions.

%
%
\bibliographystyle{plain}
\bibliography{biblio}
\end{document}